\documentclass[twocolumn]{aastex63}
\pdfoutput=1
\usepackage{amssymb}
\usepackage{amsmath}
\usepackage{verbatim}
\usepackage{graphicx}
\usepackage{dcolumn}
\usepackage{wrapfig}
\usepackage[super]{nth}
\usepackage{xcolor}
\usepackage{placeins}
\usepackage{enumitem}
\setlist[itemize]{noitemsep, topsep=0pt}

\usepackage{scalerel}
\usepackage{tikz}
\usetikzlibrary{svg.path}

\definecolor{orcidlogocol}{HTML}{A6CE39}
\tikzset{
    orcidlogo/.pic={
        \fill[orcidlogocol] svg{M256,128c0,70.7-57.3,128-128,128C57.3,256,0,198.7,0,128C0,57.3,57.3,0,128,0C198.7,0,256,57.3,256,128z};
        \fill[white] svg{M86.3,186.2H70.9V79.1h15.4v48.4V186.2z}
        svg{M108.9,79.1h41.6c39.6,0,57,28.3,57,53.6c0,27.5-21.5,53.6-56.8,53.6h-41.8V79.1z M124.3,172.4h24.5c34.9,0,42.9-26.5,42.9-39.7c0-21.5-13.7-39.7-43.7-39.7h-23.7V172.4z}
        svg{M88.7,56.8c0,5.5-4.5,10.1-10.1,10.1c-5.6,0-10.1-4.6-10.1-10.1c0-5.6,4.5-10.1,10.1-10.1C84.2,46.7,88.7,51.3,88.7,56.8z};
    }
}

\newcommand\orcidicon[1]{\href{https://orcid.org/#1}{\mbox{\scalerel*{
                \begin{tikzpicture}[yscale=-1,transform shape]
                \pic{orcidlogo};
                \end{tikzpicture}
            }{|}}}}

\usepackage{natbib}
\newcommand{\plusminus}[2]{${+{#1} \atop -{#2}}$}

\shorttitle{ZTF Microlensing Statistics} 
\shortauthors{Medford et al.}

\begin{document}

\title{Gravitational Microlensing Event Statistics for the Zwicky Transient Facility}

\author{Michael S. Medford \orcidicon{0000-0002-7226-0659}}
% \added{\correspondingauthor{Michael S. Medford} %RR
% \email{MichaelMedford@berkeley.edu}} %RR
\affiliation{University of California, Berkeley, Department of Astronomy, Berkeley, CA 94720}
\affiliation{Lawrence Livermore National Laboratory, 7000 East Ave., Livermore, CA 94550}
\affiliation{Lawrence Berkeley National Laboratory, 1 Cyclotron Rd., Berkeley, CA 94720}

\author{Jessica R. Lu \orcidicon{0000-0001-9611-0009}}
\affiliation{University of California, Berkeley, Department of Astronomy, Berkeley, CA 94720}

\author{William A. Dawson \orcidicon{0000-0003-0248-6123}}
\affiliation{Lawrence Livermore National Laboratory, 7000 East Ave., Livermore, CA 94550}

\author{Casey Y. Lam \orcidicon{0000-0002-6406-1924}}
\affiliation{University of California, Berkeley, Department of Astronomy, Berkeley, CA 94720}

\author{Nathan R. Golovich \orcidicon{0000-0003-2632-572X}}
\affiliation{Lawrence Livermore National Laboratory, 7000 East Ave., Livermore, CA 94550}

\author{Edward F. Schlafly \orcidicon{0000-0002-3569-7421}}
\affiliation{Lawrence Livermore National Laboratory, 7000 East Ave., Livermore, CA 94550}

\author{Peter Nugent \orcidicon{0000-0002-3389-0586}}
\affiliation{Lawrence Berkeley National Laboratory, 1 Cyclotron Rd., Berkeley, CA 94720}

\begin{abstract}
Microlensing surveys have discovered thousands of events with almost all events discovered within the Galactic bulge or toward the Magellanic clouds.
The Zwicky Transient Facility (ZTF), while not designed to be a microlensing campaign, is an optical time-domain survey that observes the entire northern sky every few nights including the Galactic plane. ZTF observes $\sim10^9$ stars in g-band and r-band and can significantly contribute to the observed microlensing population.
We predict that ZTF will observe $\sim$1100 microlensing events in three years of observing within $10^\circ$ degrees latitude of the Galactic plane, with $\sim$500 events in the outer Galaxy ($\ell \geq 10^\circ$).
This yield increases to $\sim$1400 ($\sim$800) events by combining every three ZTF exposures, $\sim$1800 ($\sim$900) events if ZTF observes for a total of five years, and $\sim$2400 ($\sim$1300) events for a five year survey with post-processing image stacking.
Using the microlensing modeling software \texttt{PopSyCLE}, we compare the microlensing populations in the Galactic bulge and the outer Galaxy.
We also present an analysis of the microlensing event ZTF18abhxjmj to demonstrate how to leverage these population statistics in event modeling.
ZTF will constrain Galactic structure, stellar populations, and primordial black holes through photometric microlensing.
\end{abstract}

\section{Introduction} \label{sec:introduction}

First proposed by \citet{Einstein1936LENS-LIKEFIELD}, gravitational lensing occurs when a massive object intersects the line of sight between an observer and a luminous source. The gravitational field of the intermediate object bends spacetime, acting as a lens and causing the appearance of multiple closely spaced images to an observer along this line of sight. When the massive lens and the luminous source are both stars, the multiple images of the source are separated by only microarcseconds. They are thus unresolvable and are therefore called microlensing \citep{Refsdal1964}. The photometric effect of these multiple images is an apparent amplification of the source's brightness while the source crosses behind the lens. This phenomenon is called photometric microlensing.

Microlensing possess several distinct signatures unique among astrophysical transients that aid in their discovery. If the lens and source are assumed to be point sources and the observer remains approximately stationary, the photometric light curve is a rise in brightness followed by a symmetric fall in brightness of the same timescale \citep{Paczynski1986, Paczynski1996}. This simple model is complicated by the motion of the Earth around the Sun which produces a parallax effect that perturbs the magnification depending on the time of the year that the event is observed and the location of the event in the sky \citep{Gould1992}. Microlensing is ideally achromatic; however additional sources of light in the photometric aperture, or blending, can introduce differential color changes into the transient signal \citep{DiStefano1995}. Still, images taken in multiple filters containing an approximately equal increase in brightness serve as a key piece of evidence for claiming a microlensing detection.

Observable microlensing events occur almost entirely between two stars in the Milky Way (or a nearby galaxy) as the sources and lenses rotate around the center of the galaxy. The size of the apparent ring formed by the lensed source during a theoretical perfect alignment is called the Einstein radius, given by 
\begin{equation}
\theta_E = \sqrt{\frac{4 G M_L}{c^2}\left(\frac{1}{d_L} - \frac{1}{d_S}\right)}, \label{eq:theta_E}
\end{equation}
where $M_L$ is the mass of the lens and $d_L$ and $d_S$ are the distance between the Sun and the lens and source, respectively. 
The Einstein radius is the approximate angular scale of a microlensing event in the case of a more realistic imperfect alignment between the source, lens, and observer. The centroid of the aperture's flux will perturb during a microlensing event on a scale approximately equal to the Einstein radius. This effect, known as astrometric microlensing, is extremely difficult to measure. For a typical microlensing event in the Milky Way bulge, with a source located at eight kilo-parsecs (near the center of the galaxy) and a lens halfway between the Earth and the source, a one solar mass star would produce an Einstein radius and astrometric perturbation of approximately one miliarcsecond.

The time for the luminous source to pass across the Einstein radius in the reference frame of the gravitational lens is the Einstein crossing time, given by 
\begin{equation}
    t_E = \frac{\theta_E}{\mu_\text{rel}}, \label{eq:t_E}
\end{equation}
where $\mu_{rel}$ is the relative proper motion between the source and lens as seen by the observer. This observable can be measured by fitting a photometric lightcurve with a microlensing model and identifying the timescale over which the magnification of the signal increases and then subsequently decreases. A typical microlensing event in the Milky Way bulge has an Einstein crossing time of approximately 20 days \citep{Sumi2011, Wyrzykowski2015, Mrz2017}.

Microlensing detections have resulted in many significant discoveries in the past few decades. 
Galactic models have been constrained by looking at the population statistics of microlensing events including spatial and Einstein crossing times distributions \citep{Aubourg1993, Kerins1995, Wyrzykowski2015, Navarro2020}. 
Microlensing has been used to discover and constrain exoplanet populations (\citet{Cassan2012}; See \citealt{Gaudi2012} for review) and the Nancy Grace Roman Space Telescope (formally named the Wide Field Infrared Survey Telescope) aims to significantly increase the number of exoplanets found through microlensing by $\sim$1400 \citep{WFIRST, Penny2019}. 
Looking for dark matter in the Milky Way halo using microlensing was originally proposed by \citet{Paczynski1986}, with constraints on the contribution of primordial and astrophysical black holes to the dark matter mass halo successfully executed in the years since \citep{Alcock2001, Afonso2003, Wyrzykowski2011, Niikura_2019}.
More recent work proposes detecting free floating black holes through photometric microlensing alone \citep{Lu2019}, as well as combining these observations with astrometric measurements \citep{Lu2016, Kains2016, Rybicki2018}.

Microlensing has been traditionally dominated by surveys conducted in the Galactic bulge \citep{Sumi2013, Udalski2015, Navarro2017, Kim_2018, Mrz2019} to maximize their yields, as well as the Magellanic clouds \citep{Alcock2000, Tisserand2007, Wyrzykowski2011} and M31 \citep{Novati2009, Calchi_Novati_2014} to increase the relative probability of detecting a dark matter lens relative to a stellar lens. 
The microlensing event rate is proportional to the number of luminous sources in the field of view and the mass density along the line of sight \citep{Calcino2018}, favoring pointing towards the Galactic bulge over other lines of sight in the Galaxy if attempting to maximize the microlensing event rate. 
The measurement of optical depths to microlensing by EROS-2 \citep{Hamadache2006}, optical depth and event rate by both MOA-II \citep{Sumi2013} and OGLE-IV \citep{Mrz2019}, and the study on Galactic longitude dependence by VVV \citep{Navarro2020} are all calculated in the bulge, containing fields entirely located within Galactic longitudes of $-10^\circ < \ell < 10^\circ$.
The EROS-2 spiral arm surveys \citep{Derue2001, Rahal2009} searched for microlensing at Galactic longitudes $|\ell| > 10^\circ$ but were only able to find 27 microlensing events among the 12.9 million stars observed over seven years.
Synoptic surveys (those with large footprints and wide fields of view that repeatedly observe the same fields over long stretches of time) will discover more microlensing events outside of the Galactic bulge in the outer Galaxy, and even outside of the Galactic plane, than ever before.
\citet{Sajadian_2019} predicts that the Vera C. Rubin Observatory (previously referred to as the Large Synoptic Survey Telescope) could observe anywhere from 34,000 microlensing events in its first year to 795 events per year over ten years depending on the observing strategy, showcasing the potential for an all sky survey to significantly grow the total population of microlensing events depending on the observing strategy that is implemented.
\citet{Mrz2020} lists 30 likely microlensing events discovered in the first year of the Zwicky Transient Facility's Galactic Plane Survey, and our work suggests that there remain many more events still to be discovered. Photometric filters which focus on efficiency and scale \citep{PriceWhelan2014} or introduce novel machine learning techniques that can easily scale \citep{Godines2019} may be the keys to discovering these additional events.

In this paper we present the Zwicky Transient Facility's opportunity to conduct the first all sky microlensing survey and the potential scientific contributions such a survey could enable.
In Section \ref{sec:ZTF}, we describe the Zwicky Transient Facility instrument and data.
In Section \ref{sec:uLens_estimate}, we estimate the total number of microlensing events that ZTF could discover in its first three years and methods for increasing these yields.
In Section \ref{sec:galactic_plane}, we explore the difference in population statistics for microlensing events in the outer Galaxy as compared to the Galactic bulge.
In Section \ref{sec:uLens_discovery}, we demonstrate a proof of principle for how to use the microlensing simulation software \texttt{PopSyCLE} \citep{Lam2020} to model events in the outer Galaxy and we conclude in Section \ref{sec:conclusion}.

\section{The Zwicky Transient Facility} \label{sec:ZTF}

The Zwicky Transient Facility (ZTF) is an optical time-domain survey that has been operating on the 48-inch Samuel Oschin Telescope at Palomar Observatory since March 2018 \citep{Bellm_2018}. ZTF's camera covers 47 square degrees in a single exposure, enabling coverage of the entire visible Northern sky every few nights in ZTF g-band, r-band and i-band filters with an average $2.0''$ FWHM on a plate scale of $1.01''\ \text{pixel}^{-1}$. ZTF produces a real-time alert stream triggered by transient event detections on difference images processed by the IPAC facility
\citep{Patterson2018}. In addition to these alerts, the ZTF collaboration routinely produces public data releases which contain, among other data products, lightcurves assembled from single image point spread function (PSF) photometry for every star in the northern sky which appears in a deep co-added reference image \citep{Masci2018}. Reference images are ideally constructed from 40 individual exposures resulting in an approximate r-band limiting magnitude of 22.6, although weather and visibility produces variable results.
ZTF's observing time is split between public observations (funded by the National Science Foundation's Mid-Scale Innovations Program or MSIP) and partnership observations, which are held in a proprietary period for collaboration members of the survey. The i-band filter is used only for partnership observations and is thus absent from this analysis.

\begin{figure}[htb!]
    \centering
    \includegraphics[width=0.48\textwidth, angle=0]{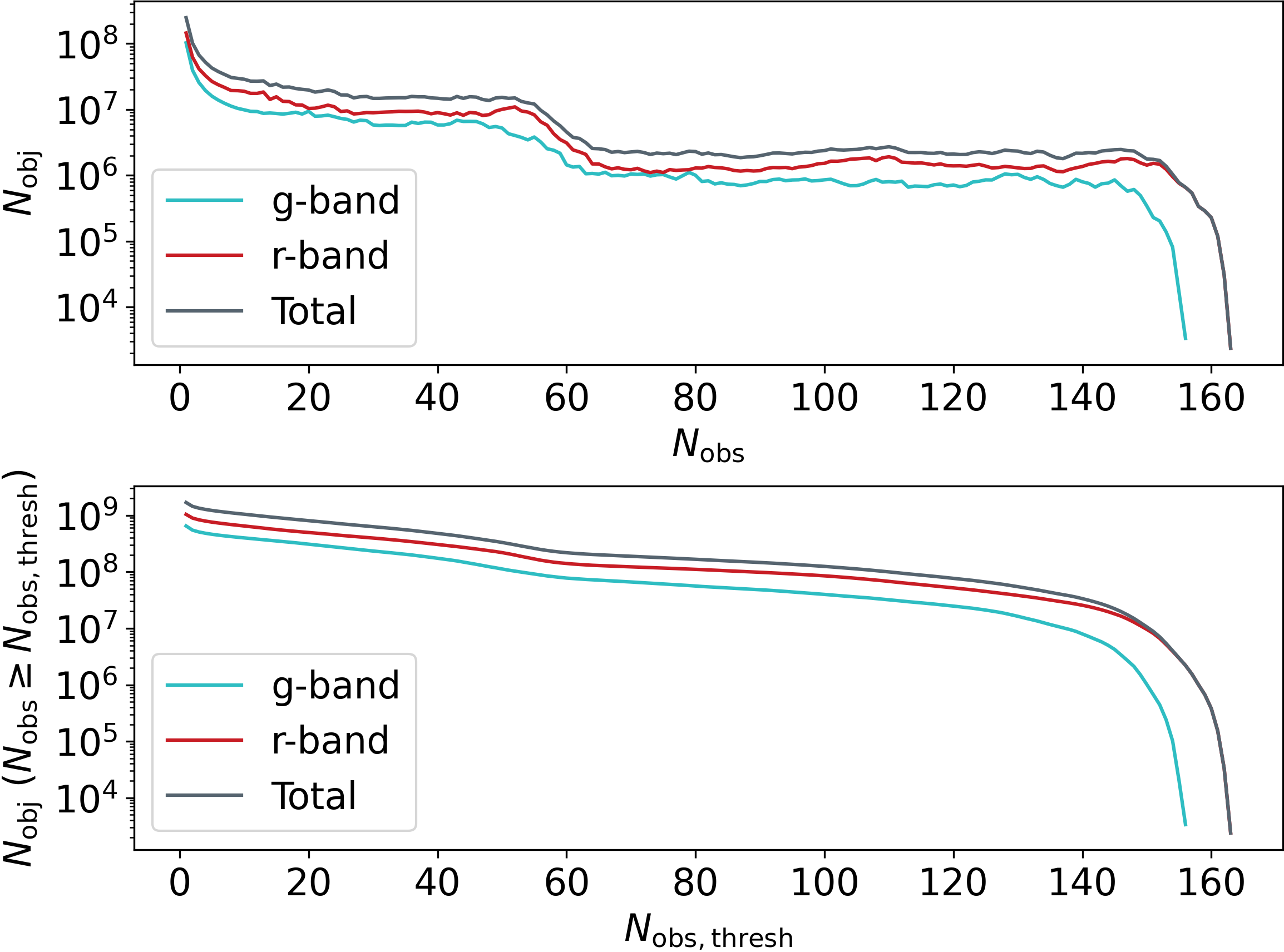}
    \caption{\small{ZTF Public Data Release 1 contains $1.7 \times 10^9$ lightcurves assembled from sources in $3.4 \times 10^6$ single-exposure images taken in g-band and r-band. Top: The number of lightcurves in each filter containing a given number of epochs, as well as the total for the two filters combined. Most lightcurves in the catalog are in fact single source detections with no subsequent observations most likely resulting from optical artifacts, moving solar system objects or faint transient sources. Bottom: The number of lightcurves with observations more than the threshold number of epochs, as well as the total for the two filters combined. For example, there are $7.8 \times 10^7$ r-band lightcurves and $1.4 \times 10^8$ g-band lightcurves with more than 60 observations, for a total of $2.2 \times 10^8$ lightcurves. Computational costs effect how many lightcurves can be searched for microlensing events and determines the minimum number of observations a lightcurve must contain. It should be noted that the ZTF data reduction pipeline treats sources detected at the same location in the sky but in the different filters as separate sources.} \label{fig:DR1_stats}}
\end{figure}

ZTF has several observing surveys covering the northern sky in r-band to a five-sigma depth of approximately $m_{lim,r} = 20.6$ and g-band to a depth of approximately $m_{lim,g} = 20.8$ every few nights \citep{Bellm_2018, Bellm2019}. 
The Northern Sky Survey observes the entire visible sky north of $-31^{\circ}$ declination in both g-band and r-band with a three night cadence and has been executed since 2018 March.
The Galactic Plane Survey \citep{Prince_2018} observes all Galactic plane fields ($-7^{\circ} < b < 7^{\circ}$) visible on a given night in both bands when the Galaxy is visible from Palomar Observatory. In total ZTF observes over 2000 square degrees in a combination of g-band, r-band, and i-band exposures every night.

The Northern Sky Survey and the Galactic Plane Survey are public surveys producing a real-time alert stream generated by transient detections on difference images \citep{Patterson_2019}. Science images of these observations are released at regular intervals, as well as a variety of data products including lightcurves assembled from single epoch photometry.  These surveys generate well sampled lightcurves for hundreds of millions of stars with non-uniform sampling due to field visibility and weather losses.
Additionally, the ZTF partnership conducts a high cadence survey in the Galactic plane with 30 second images taken on the same fields for several weeks that are released on a more infrequent basis.
All of these surveys provide excellent datasets for observing microlensing events due to short cadences and images taken in multiple filters.

On 2019 May 8, ZTF released Public Data Release 1 (DR1) containing $1.7 \times 10^9$ lightcurves assembled from sources in $3.4 \times 10^6$ single-exposure images taken in g-band and r-band for observations taken between 2018 March 17 and 2018 December 31 \href{https://www.ztf.caltech.edu/page/dr1}.
To generate these lightcurves, ZTF ran PSF photometry on both individual exposures and reference images constructed from co-adding science exposures. Sources which appeared in the reference image catalogs were used as seeds for the construction of lightcurves. Sources which appeared in the photometric catalogs of individual science images at the location of a source from the reference catalog were appended to their respective lightcurves. 
The lightcurve catalogs from DR1 contain over $8.1 \times 10^8$ lightcurves with $N_{obs} \geq 20$ from across the northern sky (Figure \ref{fig:DR1_stats}).
Both releases also include science images, reference images, subtraction images, photometric catalogs and other data products.

\section{ZTF Microlensing Estimate} \label{sec:uLens_estimate}

ZTF can be used to detect a significant number of microlensing events due to its large sky coverage, multiple filters, and repeated observations. What follows is an approximation for the number of events that ZTF could discover in its three years of operations. Here we calculate the approximate number of microlensing events ($N_\text{events}$) through combining the duration of the ZTF survey in years ($T_\text{obs}$), the number of sources ZTF observes ($N_\text{stars}^\text{ZTF}$), and the observable microlensing event rate per star per year ($\Gamma_\text{obs}$):
\begin{equation}
    N_\text{events} =\ \Gamma_\text{obs} \cdot N_\text{stars}^\text{ZTF} \cdot T_\text{obs}\ . \label{eq:N_events}
\end{equation}
The number of sources is counted from ZTF reference image photometric catalogs; however, the microlensing event rate must be estimated from simulations.

We utilize \texttt{PopSyCLE} to estimate microlensing event rates at different Galactic latitudes and longitudes. \texttt{PopSyCLE}, or Population Synthesis for Compact object Lensing Events \citep{Lam2020}, is a recently released open-source code that uses galaxy modeling and stellar population synthesis to generate realistic microlensing populations, including compact object sources and lenses. These simulations are generated along specified lines of sight in the Galaxy using stars from Galactic models \citep{Robin2003} produced by \texttt{Galaxia} \citep{Sharma2011} and compact objects determined by initial-final mass relations \citep{Kalirai2008, Sukhbold2016, Raithel2018} calculated in \texttt{PyPopStar} \citep{Hosek2020}. Estimating event rates with \texttt{PopSyCLE} provides us more physical insight into the populations of stars and compact objects undergoing microlensing than would be deduced from using analytic expressions. Simulations were run using the \texttt{PopSyCLE} v3 galaxy model (\citet{Lam2020}: Appendix A, a model which is demonstrated to accurately produce event rates in various bulge fields when compared to \citep{Mrz2019}. We note that \texttt{PopSyCLE} v3 galaxy model does not reproduce observed stellar densities in the Galactic field. However our paper adopts a relative stellar density fraction (See Section \ref{sec:N_stars_T_obs}) that corrects for this discrepancy between observed stellar densities and modelled stellar densities. This ensures that our estimate of the microlensing event rate per star are accurate.

Executing a \texttt{PopSyCLE} simulation, especially in the high stellar densities of the Galactic bulge, incurs significant computational cost and cannot therefore be performed at every ZTF field location across our estimate's footprint. The accuracy of our estimate is limited by the discrete number of simulations carried out across the Galactic plane over which we interpolate the observable event rate. Bootstrapping of the discrete simulations indicate that the precision of our event rate estimates at each location vary by approximately 10\%. The accuracy of the predicted event rate is also limited by systematic errors in the Galactic model implemented in \texttt{PopSyCLE} that we did not explore, which are known to contribute to errors in Galactic microlensing modelling \citep{Evans2002}. 

\subsection{Event Rate: $\Gamma_\text{obs}$}
The event rate in this estimate, $\Gamma_\text{obs}$, is
\begin{equation}
    \Gamma_\text{obs} = \frac{N^\text{PopSyCLE}_\text{events, detected}}{ \dfrac{N_\text{stars}^\text{ZTF}}{N_\text{stars}^\text{PopSyCLE}}\bigg\rvert_{\text{area}} \cdot T_\text{obs} \cdot N_\text{stars}^\text{PopSyCLE}} \cdot f_{\text{visibility}} . \label{eq:event_rate}
\end{equation}
The event rate is found at each sky location by dividing the number of simulated events detected $N^\text{PopSyCLE}_\text{events, detected}$ by the total number of stars in our \texttt{PopSyCLE} simulation $N_\text{stars}^\text{PopSyCLE}$ and the simulated survey duration $T_\text{obs}$. In order to account for observational effects that aren't simulated by PopSyCLE, such as blending and incompleteness in the number of stars, we then apply a correction factor ${N_\text{stars}^\text{ZTF} / N_\text{stars}^\text{PopSyCLE}}\big\rvert_{\text{area}}$ that is the ratio of stellar densities in PopSyCLE and on-sky from ZTF. This ratio is less than one across most of the Galactic plane where ZTF sees fewer stars than \texttt{PopSyCLE} due to these effects. However at the smallest galactic latitudes the ratio can be larger than one if the extinction is overestimated and there are more ZTF stars than the model predicts. However these are locations where our event rate is near zero and does not largely effect our final estimates. The rate is then corrected by a visibility completeness term $f_\text{visibility}$ that down-weights the number of microlensing events from fields proportional to their visibility by ZTF. Both the relative stellar density fraction and the visibility completeness are discussed in more detail below. We note that our predicted event rate is specifically for those events that are observable by ZTF. This would be equivalent to observational event rates reported before the completeness correction often applied by other work \citep{Sumi2013, Wyrzykowski2015, Mrz2019}.

The number of events detected ($N^\text{PopSyCLE}_\text{events, detected}$) is calculated by implementing observational cuts similar to \citet{Sumi2011, Mrz2017} in the manner outlined in \citet{Lam2020}. However \texttt{PopSyCLE}, which does not create and sample individual lightcurves, must artificially calculate some of the observational criteria of surveys.
For example, when analyzing millions of lightcurves, microlensing surveys must determine whether an increase in flux is significant.
Significant bumps in flux are with three consecutive measurements are above $3\sigma$ of the baseline flux (e.g. \citet{Mrz2017} Extended Data Table 3, \citet{Sumi2011} Table S2).
A microlensing events in \texttt{PopSyCLE} is deemed to have a significant bump in flux if 
\begin{equation*}
    F_\text{peak} - F_\text{base} > 3 \sigma_\text{base} \approx 3 \sqrt{F_\text{base}},
\end{equation*}
where $F_\text{peak}$ and $F_\text{base}$ are the peak and baseline flux, respectively.
Calculations on non-variable ZTF lightcurves of $\sqrt{F_\text{base}}$ found it to be equal to or larger than $\sigma_\text{base}$ on almost all objects, making this version of the significant bump requirement a conservative estimator.
To calculate the error on the peak and baseline flux, knowledge of the zero point magnitude $m_\text{ZP}$ is needed.
$m_\text{ZP}$ is the magnitude that corresponds to a single count in the detector $F_\text{ZP} = 1$.
Thus the flux-magnitude relation can be written
\begin{equation*}
    m - m_\text{ZP} = -2.5 \cdot \mathrm{log}_{10} (F).
\end{equation*}
$m_\text{ZP}$ is calculated for each simulated filter and the fluxes are assumed to have Poisson errors.

Table \ref{tab:mroz_cuts} contains the complete list of our selection criteria.
Both the survey duration ($T_\text{obs}\in[1, 3, 5]$ years) and minimum baseline magnitude ($19\ \text{mag}< m_\text{lim,r} < 22\ \text{mag}$) selection criteria are calculated for the stated range of values. The choice to calculate our estimate for multiple survey durations is discussed in Section \ref{sec:N_stars_T_obs}. Section \ref{sec:estimate_results} discusses applying post-processing image stacking to increase the total number of observable microlensing effects. We calculate this effect by increasing the minimum baseline magnitude accordingly. Events are required to have an Einstein crossing time, source flux fraction, and impact parameters within the limits of the stated values. The magnitude amplification $\Delta m$ is calculated by subtracting baseline magnitude from the source, lens and all neighboring stars from the magnitude at maximum amplification and must also be greater than the stated cutoff value.
All of our calculations are performed with the ZTF r-band filter by transforming \texttt{PopSyCLE}'s UBV photometry into the ZTF filter system \citep{Medford2020}. 

The observational cuts in Table \ref{tab:mroz_cuts} are chosen to result in a conservative estimate for the number of detectable microlensing events. While the average full-width half-maximum of ZTF is closer to $1.5''$, we set the seeing disk radius to the confusion limit measured in our densest fields. Setting the seeing disk radius $\theta_\text{blend} = 2.25''$ places more neighboring stars into the observational aperture and therefore increases the baseline flux of a microlensing event in a field with high stellar density. This makes the event less likely to be detected because (1) a larger baseline flux requires a larger peak flux in order to have a significant bump, (2) an event with a larger baseline flux will have smaller magnitude amplification, and (3) a larger neighbor flux decreases the source-flux-fraction. All of these effects lower the observable event rate in the Galactic bulge where more crowding occurs due to higher stellar densities.

The number of stars in the simulation ($N_\text{stars}^\text{PopSyCLE}$) results from the simulation's line of sight and the area of each simulation, which ranged from between 0.33 square degrees to 10 square degrees. There must also be a relative stellar density fraction ($N_\text{stars}^\text{ZTF} / N_\text{stars}^\text{PopSyCLE}\rvert_{\text{area}}$) applied to the number of \texttt{PopSyCLE} stars to account for blending and the discrepancies between the \texttt{PopSyCLE} Galactic model and our observations. \texttt{PopSyCLE} generates many faint stars that appear in a ZTF aperture as a single source. Failing to account for this effect would result in an artificially low event rate by over-counting the total number of observable stars. We therefore calculated the ratio of ZTF stars from reference images and \texttt{PopSyCLE} stars that overlap in the same area on the sky for each magnitude in our range of minimum baseline magnitudes.

\begin{deluxetable}{l || l}
\tabletypesize{\small}
\tablewidth{0pt}
\tablecaption{\texttt{PopSyCLE} Observational Cuts}
\tablehead{
    \colhead{Parameter/Criteria} &    
    \colhead{Value}
}
\startdata
Filter & ZTF r-band \\
Seeing disk radius, $\theta_\text{blend}$ [arcsecond] & $2.25$\\
Minimum Einstein Crossing Time, $t_E$ [days] & $\geq 3$\\
Minimum baseline magnitude, $m_\text{base}$ [mag] & $19 < m < 22$\\
Maximum impact parameter, $u_0$ & $\leq 1$\\
Removal of low-amplitude events, $\Delta m$ [mag] & $\geq 0.1$\\
Removal of highly blended events, $b_\text{sff,r}$ & $\geq 0.1$ \\
Survey duration, $T_\text{obs}$ [yrs] & 1, 3, 5\\
Significance of bump, $F_\text{peak} - F_\text{base}$ & $> 3 \sigma_\text{base}$\\
\enddata
\tablecomments0{\small{Observational cuts applied to \texttt{PopSyCLE} microlensing candidate catalogs to simulate the ZTF survey, including choosing a filter and seeing disk radius to match the instrument. Limiting magnitudes are set to a range of values to determine the effect of post-processing stacking on the final event rate. Survey durations are set to one, three and five years to measure the effect of extending the ZTF survey. See \citet{Lam2020} for more details on the implementation of each cut.}}
\label{tab:mroz_cuts}
\vspace{-1cm}
\end{deluxetable}

\begin{figure*}[htb!]
    \centering
    \includegraphics[height=12cm,angle=0]{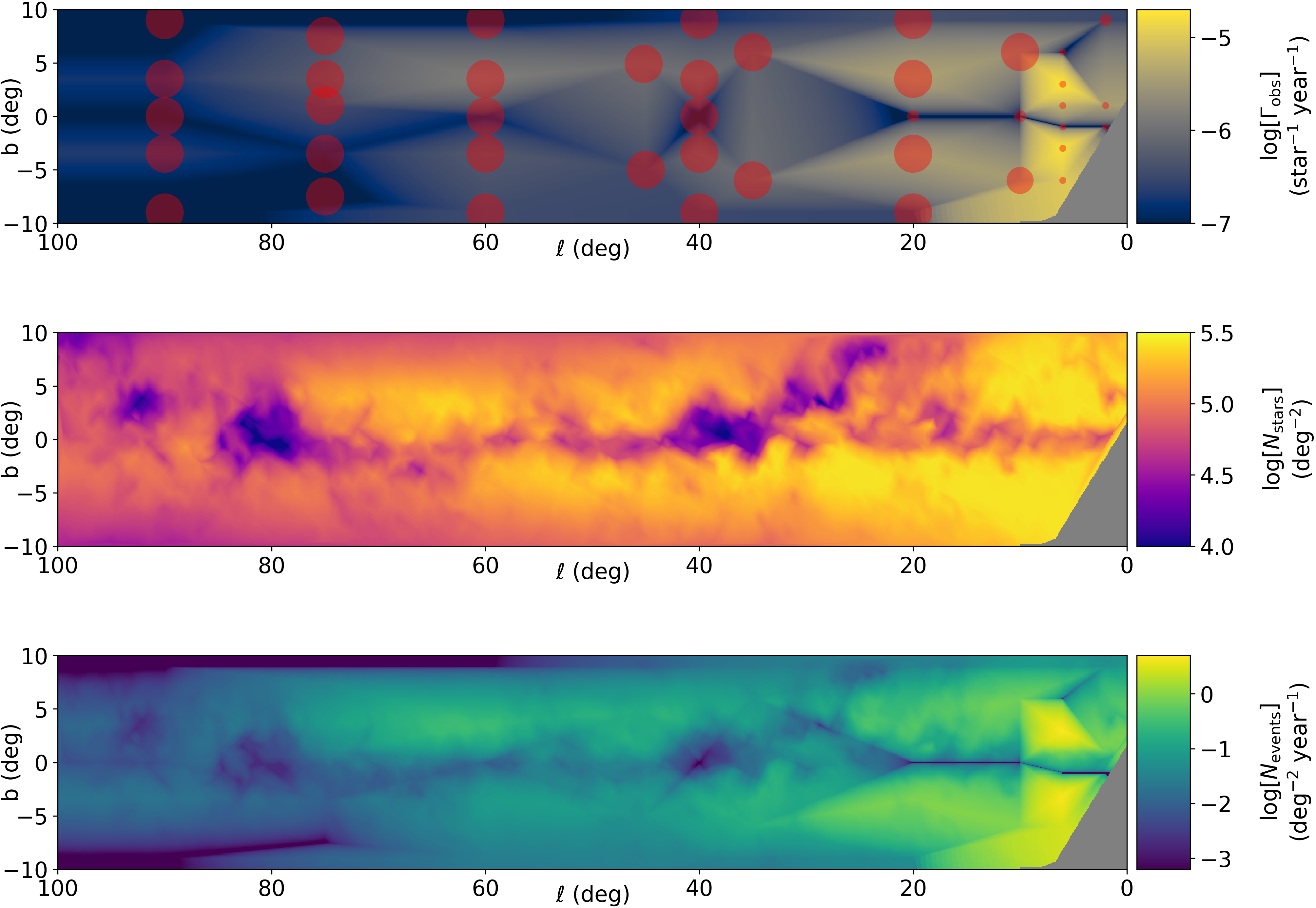}
    \caption{\small{The observable microlensing event rate density (top), stellar density (middle), and event density (bottom), for a three year survey of standard 30 second exposures. The event rate here is a detectable event rate, calculated by applying observational cuts to the \texttt{PopSyCLE} catalogs and scaling the number of sources in \texttt{PopSyCLE} to ZTF reference images. The finite grid of \texttt{PopSyCLE} runs, shown with their respective areas as red circles in the first subplot, creates lower resolution in the event rate density than the stellar density and results in an event density that maintains some of these lower resolution features. The gray area of the Galactic plane are regions which are not sufficiently visible to ZTF to render an estimate. The r-band limiting magnitude for this estimate was set at $m_{\rm lim,r} = 20.6$ magnitudes.}}
    \label{fig:uLens_estimate_on_sky}
\end{figure*}

One might note that the number of ZTF stars ($N_\text{stars}^\text{ZTF}$) and the number of \texttt{PopSyCLE} stars ($N_\text{stars}^\text{PopSyCLE}$) both appear twice in Equations \ref{eq:N_events} and \ref{eq:event_rate} and conclude that these terms can both be cancelled. If simulations were able to be carried out at all locations across the Galactic plane this would be correct because the number of events detected ($N_\text{events, detected}^\text{PopSyCLE}$) is itself an accurate measure of the number of events ZTF could detect toward that line of sight. However our strategy of constructing an interpolated grid of event rates requires that we convert the number of events detected into a rate per star. This allows us to multiply the interpolated event rate density (star$^{-1}$ year$^{-1}$) by the stellar density (deg$^{-2}$) to calculate the event density (year$^{-1}$ deg$^{-2}$).

The visibility completeness ($f_\text{visibility}$) is determined for each field by simulating observation of that field throughout the year and calculating the fraction of nights per year that the field is visible for more than 30 minutes at an airmass less than 2.1. The event rate for a field is down-weighted by this fraction because only events that are observed during peak would be detected as microlensing events. The ZTF Northern Sky Survey and Galactic Plane survey ensure that a Galactic plane field that is visible will be observed and therefore this simulated fraction accurately represents the relative fraction of microlensing events that will be observed to peak within the survey duration of ZTF.

\subsection{Number of Stars: $N_\text{stars}^\text{ZTF}$, Survey Duration: $T_\text{obs}$}
\label{sec:N_stars_T_obs}
The ZTF DR1 contains reference images photometric catalogs constructed from deep co-additions. We count the number of sources in each field, using the range of minimum baseline magnitudes as a limiting magnitude cut on the catalog. For each of these magnitude cuts, we generate an interpolated stellar density map.

ZTF has a planned operation timeline of three years with almost two years of operations already completed. Longer surveys are able to observe events with longer Einstein crossing times, creating a non-linear increase in the number of observable events with increasing survey duration. Our estimate was performed with a $T_\text{obs}$ equal to one, three, and five years in order to demonstrate the increased yields in future ZTF data releases, as well as the benefit of continuing operations beyond the planned operation timeline. The \texttt{PopSyCLE} simulated survey duration was set to the same time in order to remove long duration microlensing events from the observable event rate that could not be detected in the duration of the survey.

\begin{figure}[b]\setlength{\hfuzz}{0.9\columnwidth}
\begin{minipage}{\textwidth}
    \centering
    \includegraphics[height=6.1cm]{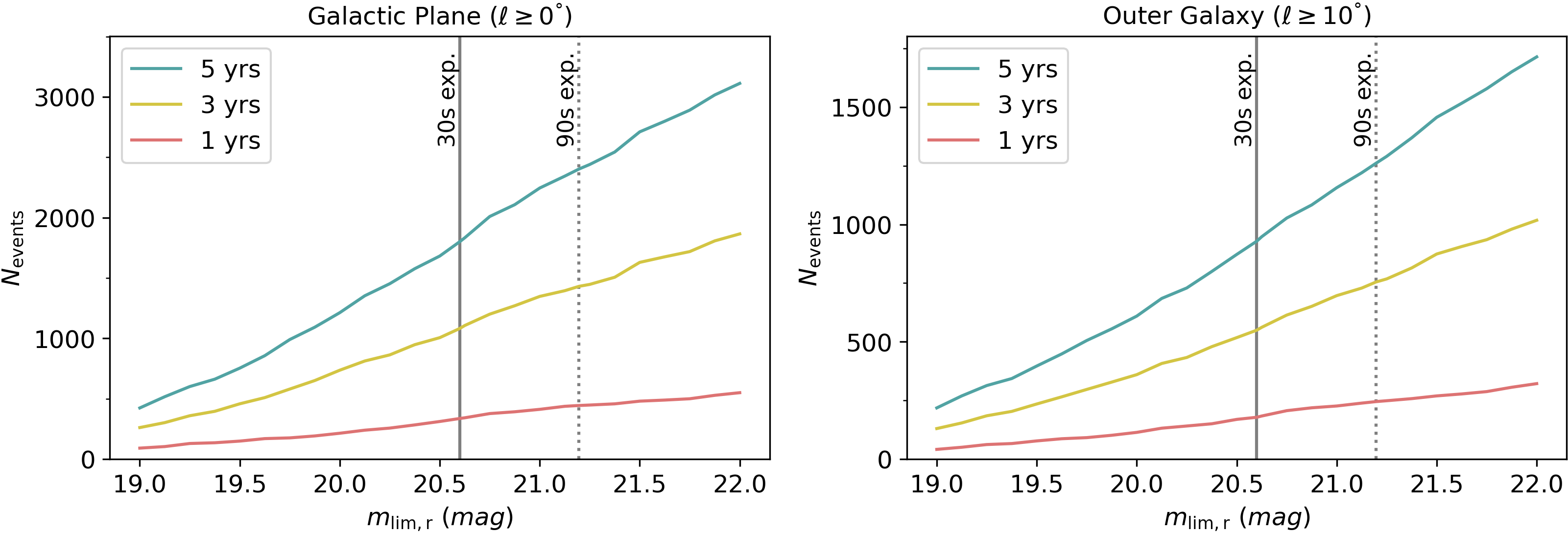}
    \caption{\small{The total number of microlensing events observable by ZTF at different limiting magnitudes for one year (red), three years (yellow), and five years (blue) in the visible Galactic plane (top) and the outer Galaxy (bottom). ZTF will observe $\sim$1100 events over three years of operation at a r-band limiting magnitude of 20.6 (vertical black), with $\sim$500 of these events occurring in the outer Galaxy ($\ell \geq 10^\circ$). If every three images are stacked together before generating photometric catalogs, the limiting magnitude would increase to 21.2 magnitudes (vertical dashed black) and would increase the yield to $\sim$1400 events over three years, with $\sim$800 events in the outer Galaxy. This stacking strategy would result in a cadence of three to five days. The total number of events observed would increase to $\sim$2400 if the ZTF survey were extended to five years and this image stacking procedure were implemented, with $\sim$1300 events in the outer Galaxy.}}
    \label{fig:uLens_estimate_mag_cut}
\end{minipage}
\end{figure}

\subsection{Interpolation Across the Galactic Plane}
We ran \texttt{PopSyCLE} simulations and calculated stellar counts from ZTF reference images for fields in the Galactic plane visible to ZTF, at galactic longitudes from $100^{\circ} > \ell > 0^{\circ}$ and galactic latitudes from $-10^{\circ} < b < 10^{\circ}$ (Figure \ref{fig:uLens_estimate_on_sky}). Preliminary investigation suggested that extending the search to $|b| > 10^{\circ}$ and $\ell > 100^{\circ}$ would not significantly increase the predicted yield of microlensing events, although ZTF will observe these fields. The locations of our \texttt{PopSyCLE} simulations roughly cover the morphology of the Galactic plane and were used to create a linear interpolation of the event rate density (star$^{-1}$ year$^{-1}$) and stellar density (deg$^{-2}$). \texttt{PopSyCLE} simulations were run at different sizes depending on their sky location in order to strike a balance between computational runtime and statistically significant numbers of microlensing events. Simulations away from the Galactic bulge were run on patches ranging from 1 deg$^{2}$ to 10 deg$^{2}$, making the observable microlensing event rates at these locations an average over the simulation's field of view. Simulations in the Galactic bulge where executed with an area of 0.33 deg$^{2}$. Interpolating over the Galactic plane required choosing a scheme that accurately reflected the dynamic range of the stellar density, which we expect to be an approximate tracer of the event rate. We therefore choose to apply a linear interpolation and nearest extrapolation to our grid of event rates. Our sparse sampling is subject to interpolation errors that could effect our final results by up to a factor of two. 

The location of Mount Palomar in the northern hemisphere limits the visibility to fields in the Galactic bulge closest to the Galactic center. The lack of data in these fields prevents us from making a measurement of the number of stars because too few exposures were taken in these fields to generate reference images. However, individual images of these fields have been taken by ZTF and some of fields are expected to have reference images by the end of the telescope's three year lifespan. Microlensing predictions and searched can be recalculated after the completion of ZTF to increase their accuracy and yields.

\vspace{-1mm}

\subsection{Results: ZTF Microlensing Event Statistics}
\label{sec:estimate_results}

\begin{table*}[t]
  \centering
  \caption{Description of Fiducial Microlensing Simulations}
    \begin{tabular}{c || ccc}
      \hline
      \hline
      Property & Inner Galactic Bulge & Outer Galactic Bulge & Outer Galaxy \\
      \hline
      Galactic Longitude $\ell$ & $2.0^\circ$ & $6.0^\circ$ & $45.2^\circ$\\
      Galactic Latitude $b$ & $1.0^\circ$ & $3.0^\circ$ & $4.9^\circ$ \\
      \texttt{PopSyCLE} Area & 0.33 deg$^2$ & 0.33 deg$^2$ & 10 deg$^2$ \\
      \texttt{PopSyCLE} Extinction in ZTF r-band at 8 kpc & 6.6 mag & 2.4 mag & 1.8 mag \\
      ZTF Stellar Density at $m_\text{lim,r} = 20.6$ mags & $2.76 \times 10^7$ deg$^{-2}$ & $5.01 \times 10^6$ deg$^{-2}$ & $1.76 \times 10^5$ deg$^{-2}$\\
    \hline
    \end{tabular}
    \caption*{\small{The three fields were chosen to demonstrate the differences in microlensing populations between the Galactic bulge and the outer Galaxy. The Galactic bulge fields represent the range of typical observations in the bulge with significantly higher stellar densities and extinctions than a field in the outer Galaxy. The Galactic bulge fields are smaller in order to be computationally tractable, while the outer Galaxy field is larger to generate a statistically significant numbers of microlensing events.}}
  \label{tab:simulation_descriptions}
\end{table*}

ZTF will observe $\sim$1100 events over its fiducial three years of operation, assuming an r-band $5\sigma$ limiting magnitude of $m_\text{lim,r} = 20.6$ (Figure \ref{fig:uLens_estimate_mag_cut}). $\sim$600 events occur in the Galactic bulge ($\ell < 10^{\circ}$) where both the event rate and stellar density are large. This appears to validate \newpage \noindent the observing strategy taken by most microlensing campaigns to observe in the Galactic bulge where the event rate is highest. However $\sim$500 events occur throughout the outer Galaxy ($\ell > 10^{\circ}$) despite the drop-off in event rate and stellar density at larger Galactic longitudes. This is driven by the increased efficiency at detecting events further out in the plane \citep{Sajadian_2019} where reductions in crowding and consequently less confusion from neighboring stars in the photometric aperture make it easier to detect events relative to the bulge. The yields in the outer Galaxy are also increased due to the much larger footprint it covers compared to the bulge. Few microlensing events have been found at such large Galactic latitudes \citep{Nucita2018, Dong2019, Wyrzykowski2020}. Extending the survey duration of ZTF to five years would increase the number of detectable events to $\sim$1800, with $\sim$900 events occurring in the outer Galaxy. Increasing the lifetime of the survey captures more of the long duration events particularly present at the larger Galactic longitudes, as well as increasing the number of short duration events across the entire plane.

The ZTF Northern Sky Survey and Galactic Plane Survey take 30 second exposures with a cadence of one to three days across the Galactic plane. The shift in the distribution of Einstein crossing times discussed in Section \ref{sec:galactic_plane} means that most microlensing events would still be observable with a cadence of three to five days. Surveys such as the ZTF Uniform Depth Survey (Goldstein et. al. in prep) are creating photometric catalogs from co-additions of science images that simulate a deeper and longer cadence survey than ZTF's current operations. Combining every three observations on ZTF would increase the r-band limiting magnitude to 21.2 magnitudes, increasing the three year yield to $\sim$1400 events ($\sim$800 events in the outer Galaxy), with $\sim$2400 ($\sim$1300) microlensing events observable if ZTF were extended to five years.

We stress here that the majority of these microlensing events will occur outside of the Galactic bulge and therefore beyond the footprint of most previously conducted microlensing campaigns. This presents the opportunity to constrain Galactic models and measure stellar population statistics in ways previously not possible with gravitational microlensing. While our method does not make extremely precise predictions, it does demonstrate that executing a microlensing survey with ZTF will yield significant numbers of microlensing events through the less explored Galactic plane.

% \newpage 
\section{Microlensing Population Properties in the Outer Galaxy ( $\ell \geq 10^\circ$)} \label{sec:galactic_plane}

Simulations of microlensing generated by \texttt{PopSyCLE} at these larger Galactic longitudes predict significant differences in the population distributions as compared to microlensing events the Galactic bulge. To highlight some of the difference in the microlensing populations at these different locations, we selected several fiducial fields to compare against each other. Analysis was performed in (1) the inner Galactic bulge, (2) the outer Galactic bulge, and (3) the outer Galaxy. Details of the characteristics of these fields can be found in Table \ref{tab:simulation_descriptions}. These fields are not meant to serve as representative of the Galactic bulge or outer Galaxy in their entirety, but were chosen in order to highlight the significant differences that can be found between the microlensing populations at different locations in the Galaxy. Such differences must be examined in order to properly model microlensing events and measure the physical parameters of a microlensing event. We demonstrate these effects on modeling in Section \ref{sec:uLens_discovery}.

\begin{figure*}[htb!]
\centering
\begin{minipage}[b]{.46\textwidth}
    \includegraphics[width=\textwidth]{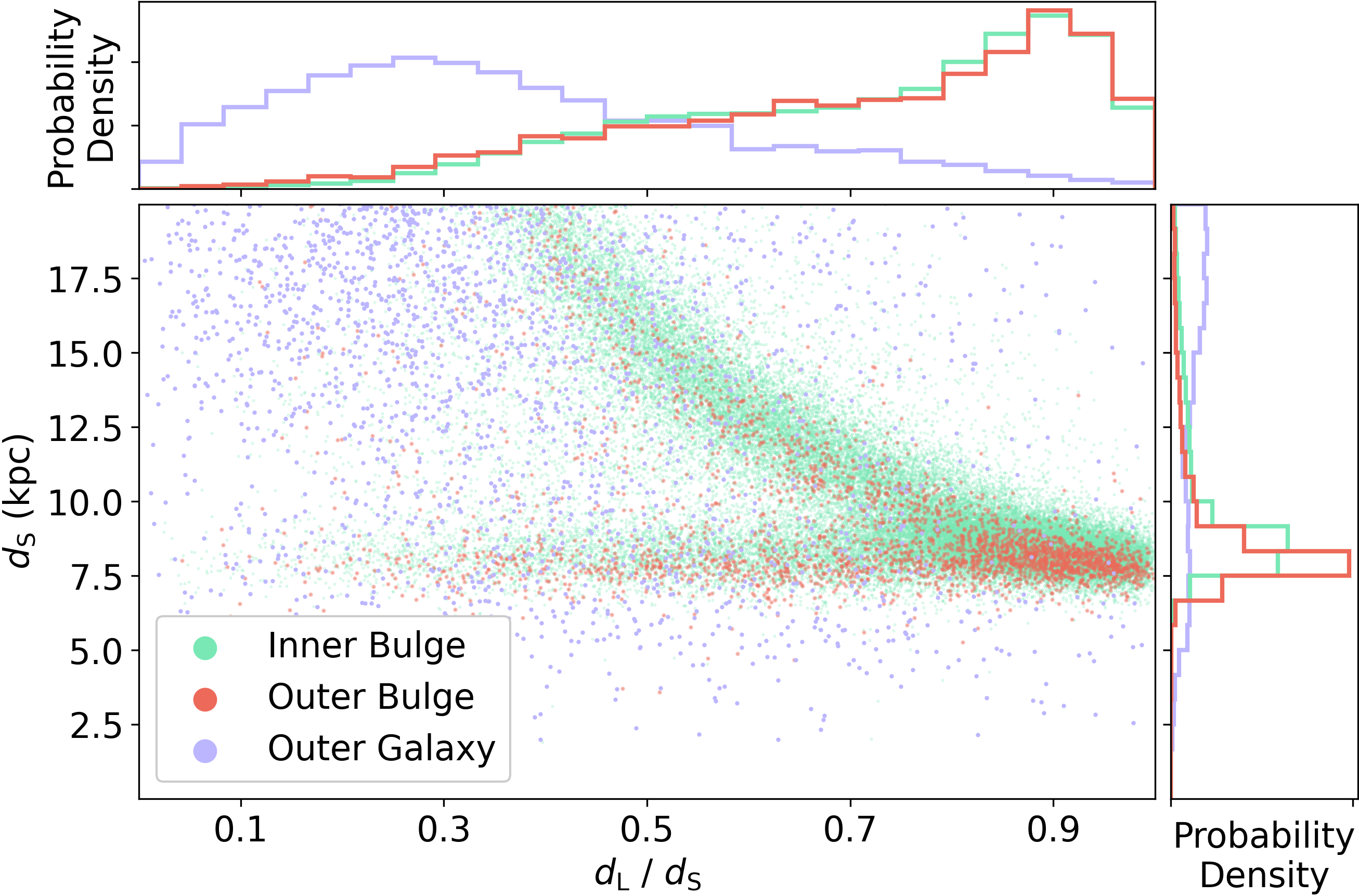}
    \caption{\small{The location of microlensing sources in the galaxy against their distance ratio in the inner bulge (green), outer bulge (red) and in the outer Galaxy (purple), with histograms on both axes showing the marginalized distributions of the parameters. Events in the direction of the Galactic bulge have lenses and sources almost entirely located in the bulge ($\sim$6-11 kiloparsecs away). The outer Galaxy events are more evenly spread out in source distance, with an approximately linear increase in sources at further distances. This results in an overall lower average distance ratio that must be appropriately used as a prior for any microlensing modeling in the outer Galaxy.\label{fig:dist_ratio}}}
\end{minipage}\qquad
\begin{minipage}[b]{.46\textwidth}
        \includegraphics[width=\textwidth, angle=0]{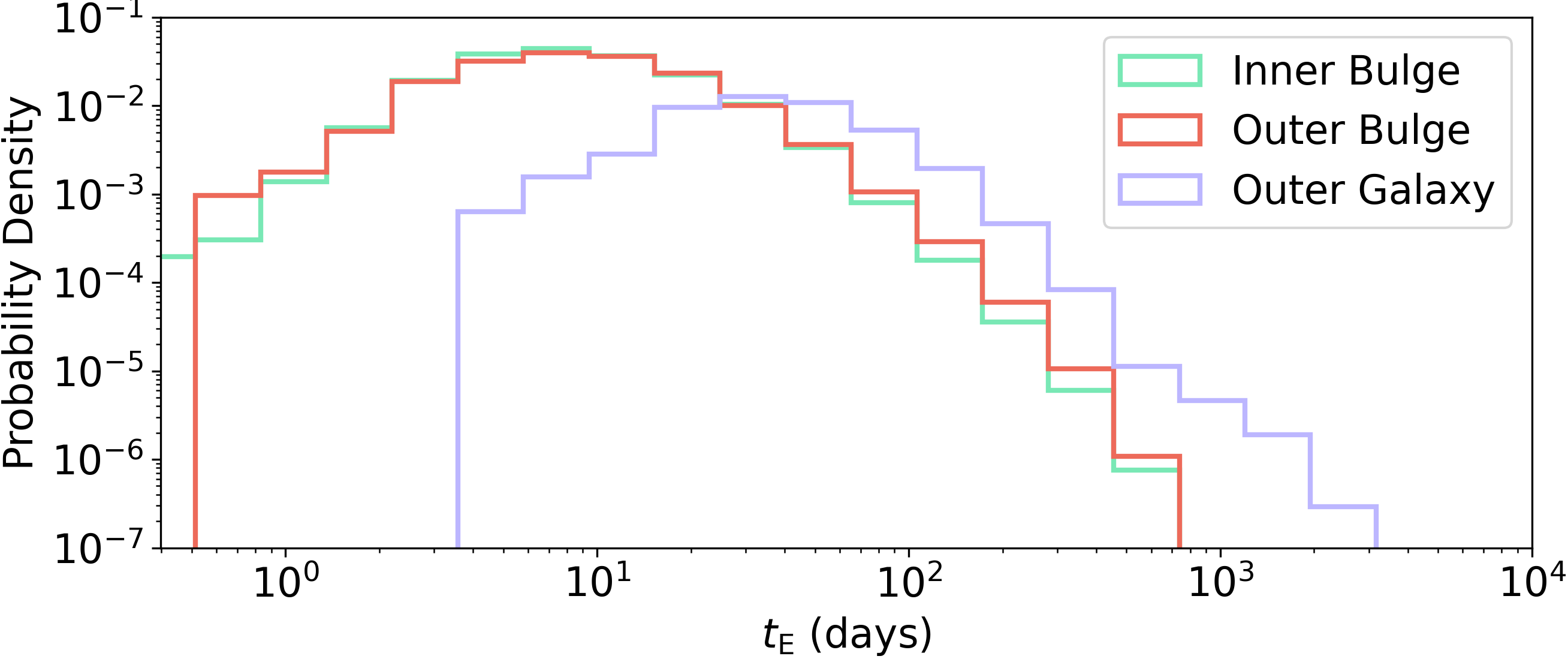}
        \caption{\small{The distribution of the Einstein crossing time in the inner bulge (green), outer bulge (red) and in the outer Galaxy (purple), with histograms on both axes showing the marginalized distributions of the parameters. Both Galactic bulge fields have an average Einstein crossing time of approximately 25 days, in alignment with previous work. However the outer Galaxy distribution averages around 80 days and stretches out beyond 1000 days in far excess of the Galactic bulge fields, with almost no events having an Einstein crossing time shorter than 10 days. Surveys can afford a longer observational cadence when searching for microlensing in the outer Galaxy due to this shift in the Einstein crossing time distribution.}\label{fig:einstein_crossing_time}}
\end{minipage}
\end{figure*}

\begin{figure*}[htb!]
\centering
\begin{minipage}[b]{.46\textwidth}
        \includegraphics[width=\textwidth, angle=0]{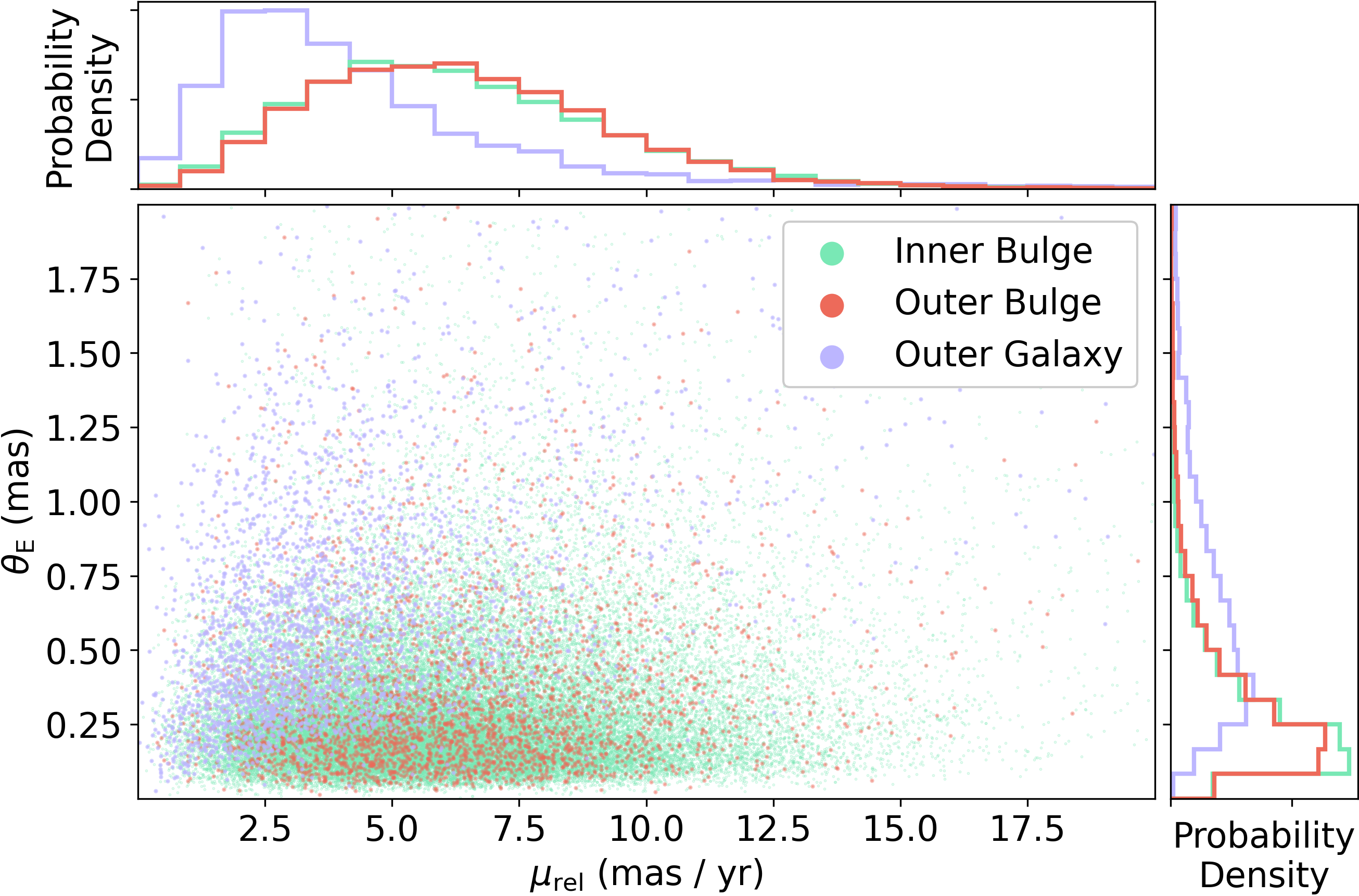}
        \caption{\small{The size of the Einstein lens radii against the relative proper motions between the sources and lenses in the inner bulge (green), outer bulge (red) and in the outer Galaxy (purple), with histograms on both axes showing the marginalized distributions of the parameters. Microlensing events in the outer Galaxy have longer Einstein crossing times than those in the bulge due to their shorter relative proper motions and larger Einstein radii. The increased Einstein radii of outer Galaxy events makes them easier to follow up astrometrically in order to break the microlensing mass-distance degeneracy. However their slower relative proper motions results in a longer time before sources and lenses are resolvable on the sky due to separation.} \label{fig:prop_motion}}
\end{minipage}\qquad
\begin{minipage}[b]{.46\textwidth}
      \includegraphics[width=\textwidth, angle=0]{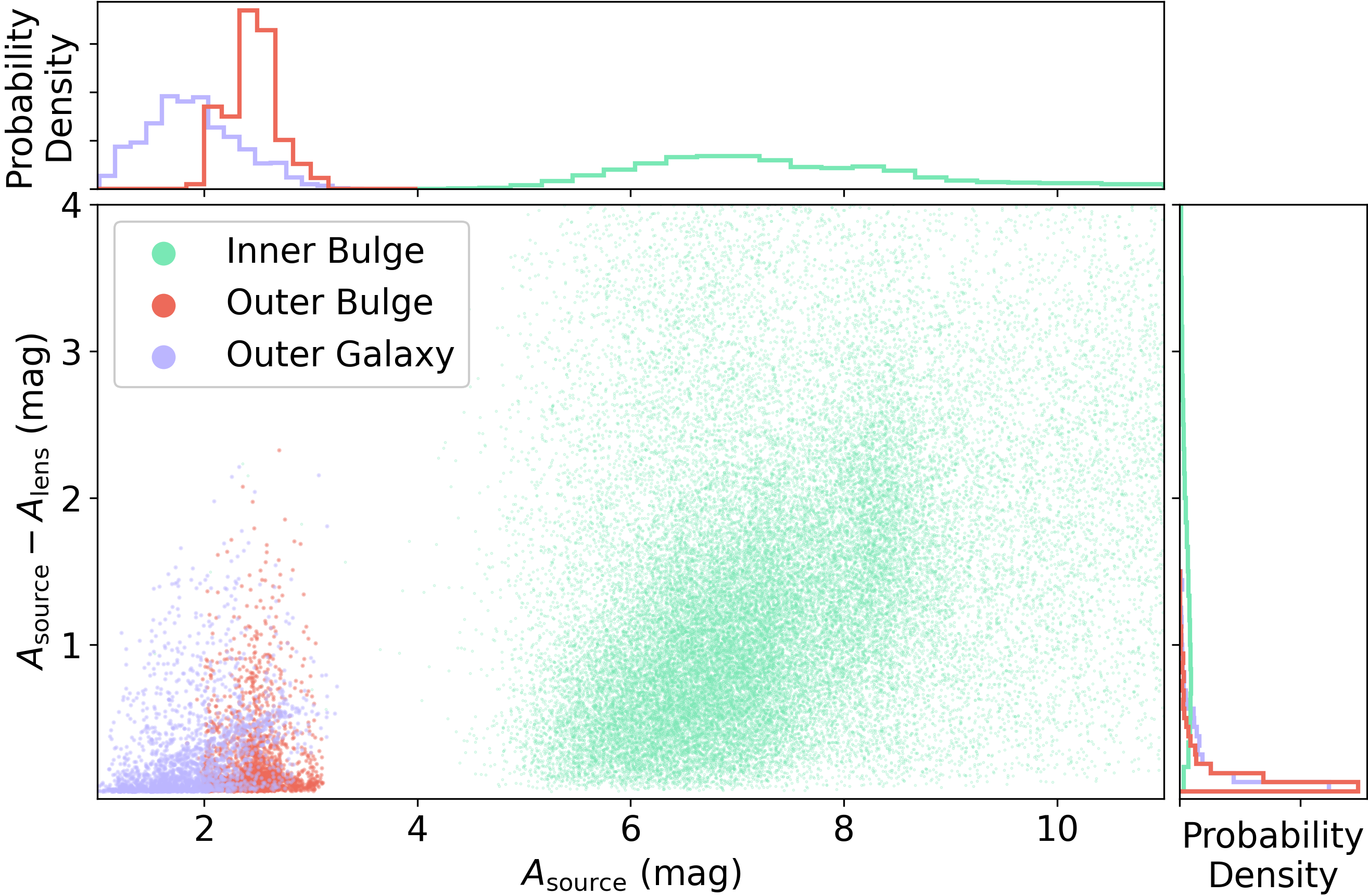}
        \caption{\small{Extinction in the r-band to the sources and lenses of microlensing events in the inner bulge (green), outer bulge (red) and in the outer Galaxy (purple). Extinction towards the inner  bulge is much larger than towards the outer Galaxy and even toward the outer bulge. This makes estimating the difference in extinction toward microlensing sources and lenses much more difficult in the inner bulge (up to 4 magnitudes) than in the outer bulge and outer Galaxy (less than 1 magnitude), despite the tighter constraints on both the distance to the sources and the distance ratio in the bulge (see Figure \ref{fig:dist_ratio}).} \label{fig:extinction}}
\end{minipage}
\vspace{1cm}
\end{figure*}

\subsection{Distance Ratio}
The distance ratio between the lenses and sources ($d_L / d_S$) is largely determined by the mass density along a given line of sight in the galaxy. Therefore it should not be surprising that the distribution of the distance ratio is different along different lines of sight (Figure \ref{fig:dist_ratio}). The average distance ratio towards the Galactic bulge is approximately 0.8, with sources in the inner bulge appearing at slightly larger distances. The distribution of distances to sources and lenses toward the outer Galaxy is significantly different, with a distance ratio peaking at approximately 0.25. The difference in these two distributions is driven by the different distance distributions of both the sources and the lenses. Sources and lenses towards the Galactic bulge are almost entirely located in the bulge ($\sim$6-11 kiloparsecs away), while the number of sources in the outer Galaxy increases approximately linearly at further distances.

\subsection{Einstein Crossing Time, Einstein Radius, and Relative Proper Motion} \label{sec:einstein_radius}
A commonly noted difference between microlensing populations in the Galactic bulge and the outer Galaxy is the distribution of Einstein crossing times \citep{Sajadian_2019} and the trend toward longer Einstein crossing times at larger Galactic longitudes \citep{Mrz2019}. We find a similar trend, with lines of sight further out along the outer Galaxy having larger crossing times (Figure \ref{fig:einstein_crossing_time}), averaging approximately 25 days in the bulge and almost 80 days in the outer Galaxy. This divergence is driven by the difference in relative proper motions and the Einstein radii between the two populations (Figure \ref{fig:prop_motion}). The events in the bulge have mostly small Einstein radii and large relative proper motions, both pushing the Einstein crossing time toward smaller values (Equation \ref{eq:t_E}). The opposite is found in the outer Galaxy, where lenses with large Einstein radii are crossed by luminous sources at relatively slower speeds.

Events in the Galactic bulge are difficult to measure astrometrically due to their smaller Einstein radii caused by the relatively similar distances to their sources and lenses as compared to the outer Galaxy (Equation \ref{eq:theta_E}). However microlensing events in the outer Galaxy will be easier to measure astrometrically due to their larger Einstein radii, with a significant number of events having radii larger than one miliarcsecond. Astrometric measurement is a key method for breaking the mass-distance degeneracy that often plagues microlensing modeling. Lens masses will be better able to be constrained in the galactic plane because of these larger Einstein radii. It should also be noted that the decrease in relative proper motion will make it harder to observe these events with high resolution follow-up, which can determine the contribution to the aperture flux originating from neighbors and possibly observe source-lens separation after long periods of time.

\begin{figure}[!htb]
    \centering
    \includegraphics[width=0.45\textwidth, angle=0]{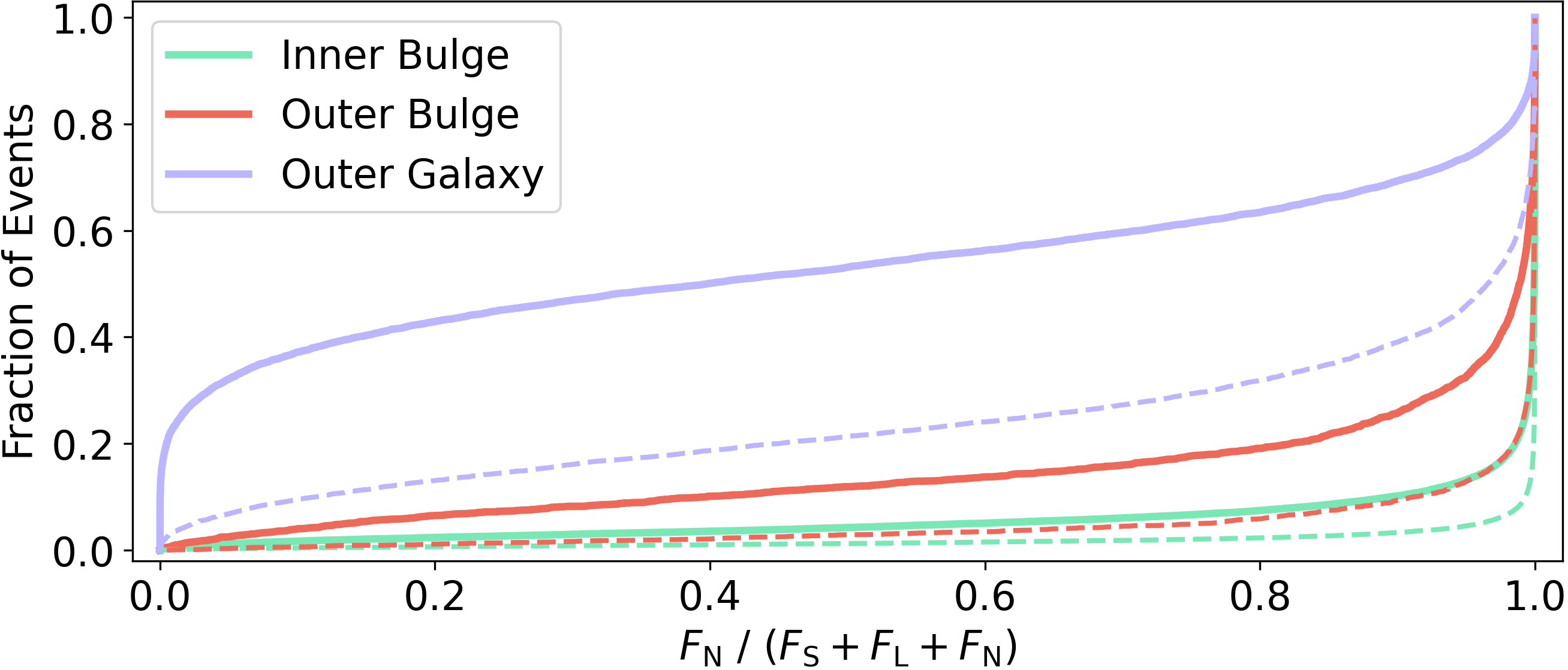}
    \caption{\small{Fractional contribution of the flux from neighboring stars in a $\theta_\text{blend} = 1.0''$ observational aperture (solid) for the inner bulge (green), outer bulge (red) and outer Galaxy (purple), as well as a larger $\theta_\text{blend} = 2.25''$ aperture (dashed) for the outer Galaxy. Increasing the size of the observational aperture has a small effect on bulge fields where even the smaller aperture is dominated by the presence of neighbor flux. However improved seeing conditions in the outer Galaxy minimizes the contamination from neighbor flux to microlensing events, making these events easier to model.} \label{fig:f_neighbors}}
\end{figure}

\subsection{Extinction}
In order to infer the absolute magnitude and therefore spectral type of a microlensing source and lens we require an estimate of the extinction to both. This is difficult in the inner bulge due to large amounts of extinction that can be significantly different between the source and the lens. \texttt{PopSyCLE} uses the color excess values from the \citet{Schlegel1998} 3-D dust maps and the \citet{Damineli2016} reddening law to calculate interstellar extinction.
\citet{Lam2020} Appendix B outlines how this results in accurate magnitudes and colors for stars throughout the bulge and greater Galactic plane.

Figure \ref{fig:extinction} shows the r-band extinction to sources and lenses in our three fields with significantly more extinction occurring in the inner Galactic bulge than the other fields as is expected. Extinction toward the inner Galactic bulge varies between five and nine magnitudes, with sources and lens having a difference of zero to four magnitudes despite their relatively equal distances. The outer Galactic bulge and outer Galaxy fields are more similar, each having less than three magnitudes of extinction to their sources and averaging approximately 0.1 magnitudes difference between the source and lens. We use the tightness of this distribution in the outer Galaxy in our estimate of the source and lens stellar types in Section \ref{sec:uLens_discovery}.

\subsection{Contribution of Neighbors to Blended Light}

The source flux fraction, $b_\text{sff}$, is 
\begin{align*}
    b_\text{sff} = \frac{f_\text{S}}{(f_\text{S} + f_\text{L} + f_\text{N})},
\end{align*}
or the flux from the source $f_\text{S}$ divided by the sum of the fluxes from the source, lens $f_\text{L}$ and any neighbors that reside within the observational PSF $f_\text{N}$. The source flux fraction is often dominated by the presence of neighbors (stars that fall in the aperture but are neither the source nor the lens) in crowded Galactic fields. Figure \ref{fig:f_neighbors} shows the contribution of flux from neighboring stars in an aperture of radius of $\theta_\text{blend} = 1.0''$ to simulate high quality seeing conditions on ZTF and an aperture of radius $\theta_\text{blend} = 2.25''$ to match the conservative estimate used throughout this analysis.
Decreasing the observational aperture and the surrounding stellar density both reduce the fraction of flux originating from neighbors. Over half of the events have in excess of 99\% of their flux originating from neighbors in all fields observed with the larger aperture.
This causes the source flux fraction to shift towards zero in these microlensing populations. However observing the outer Galaxy field with a smaller aperture results in half of the microlensing events having less than 40\% neighbor flux.
This makes modelling the source flux fraction of microlensing events along these lines of sight easier because one can reasonably use a strong prior that assumes a only a small amount of neighbor flux present, assuming that the event has been observed with high quality photometry.
The population of events in the outer Galaxy is almost entirely devoid of neighbor flux due to lower stellar densities. This makes modelling the source flux fraction of microlensing events along these lines of sight easier because one can reasonably use a strong prior that assumes little to no neighbor flux present.
 
\subsection{Implications for Outer Galaxy Microlensing}
Future microlensing searches with ZTF must consider how the distribution of microlensing parameters across the outer Galaxy differs from those distributions in the Galactic bulge. While the shift in Einstein crossing times to larger values at these Galactic longitudes have been predicted, other microlensing parameters also change at these plane locations and must be considered to properly model events, measure properties of stellar populations and constrain galactic structure.

Modeling microlensing events with a Bayesian analysis requires selecting priors that are physically motivated by population statistics. The differences between the statistics of  Galactic bulge and outer Galaxy populations should be noted as both an opportunity and a warning. Priors for microlensing populations that are appropriate for the Galactic bulge cannot be extended to analysis conducted in the outer Galaxy, and instead probabilistic priors should be derived from microlensing simulations performed at the location of microlensing events. We have made the catalogs of our fiducial microlensing populations available for public download at \url{https://portal.nersc.gov/project/uLens/Galactic_Microlensing_Distributions/} following the data structure outlined in the \texttt{PopSyCLE} documentation. Future work will include releasing the full set of catalogs generated by our grid of \texttt{PopSyCLE} simulations.

\section{Example Outer Galaxy Microlensing Event Analysis} \label{sec:uLens_discovery}

The different microlensing population distributions in the outer Galaxy open the door to new opportunities for how to fit microlensing events. We here present an example ZTF microlensing event analysis to demonstrate how modelling outer Galaxy microlensing events can take advantage of these population statistics.

\subsection{Event Selection}

\citet{PriceWhelan2014} investigates statistical methods for detecting microlensing events in non-uniformly spaced time domain surveys that cover large areas of the sky. The heterogeneous time sampling and increased number of lightcurves in such a survey makes it challenging to adapt detection methods optimized from searches in the Galactic bulge to searches across the outer Galaxy. A method for finding microlensing events in surveys with a larger footprint must be extremely inexpensive to calculate for each lightcurve in order to scale efficiently. \citet{PriceWhelan2014} concludes that the von Neumann ratio (the mean square successive difference divided by the sample variance) works well as a statistic for filtering microlensing events that is inexpensive enough to be calculated for many lightcurves while discerning enough to avoid many of the false positives that other statistics routinely produce.

We calculated the von Neumann ratio on all lightcurves in the ZTF DR1 with $N_\text{obs} \geq 100$, totalling approximately $1.25 \times 10^8$ lightcurves. We removed all lightcurves with more than one cluster of consecutive observations more than $3\sigma$ above the median brightness of the source. This left 136,638 lightcurves in our sample. We selected the 2\% of lightcurves with the largest von Neumann ratios and matched sources with both g-band and r-band lightcurves at the same sky location. 28 objects appeared to have amplification in the lightcurves of both filters which was achromatic to within approximately 0.5 magnitudes. However 25 of the objects had amplification that was quasi-periodic or slowly rising in what appeared by eye unlikely to be microlensing. Those lightcurves with a characteristic microlensing shape were fit by microlensing models, resulting in one microlensing detection.

The purpose of this search and analysis was to verify that current ZTF cadence and filter coverage is capable of observing a measurable microlensing event. We emphasize that this process was meant to serve as neither a complete search nor a scalable model for microlensing discovery. \citet{PriceWhelan2014} outline a sophisticated statistical approach for determining cuts on statistical parameters, such as the von Neumann ratio, that are tailored to finding microlensing events. Our efforts were not to replicate this procedure but to instead scan the DR1 dataset using one of these statistics until a microlensing event was found. Our focus was on finding an example microlensing event to demonstrate how to use \texttt{PopSycle} to improve microlensing modeling, not to demonstrate a method for microlensing discovery. An improved search strategy could follow the detection algorithm of \citet{PriceWhelan2014} and include (1) removing lightcurves not simply by the number of observations but on the quality of those observations, (2) cutting lightcurves on a von Neumann ratio threshold determined from injecting artificial microlensing events into lightcurves to determine a false positive rate, (3) recalculating the von Neumann ratio after subtracting off a microlensing model, and more. Our search included none of these steps and we are therefore not surprised to find such a small completeness. Future work will include implementing a robust microlensing discovery algorithm resulting in measurements of the microlensing optical depth and event rate across the ZTF footprint.

\subsection{Event Analysis}

\begin{figure}[!htb]
    \centering
    \includegraphics[width=0.45\textwidth, angle=0]{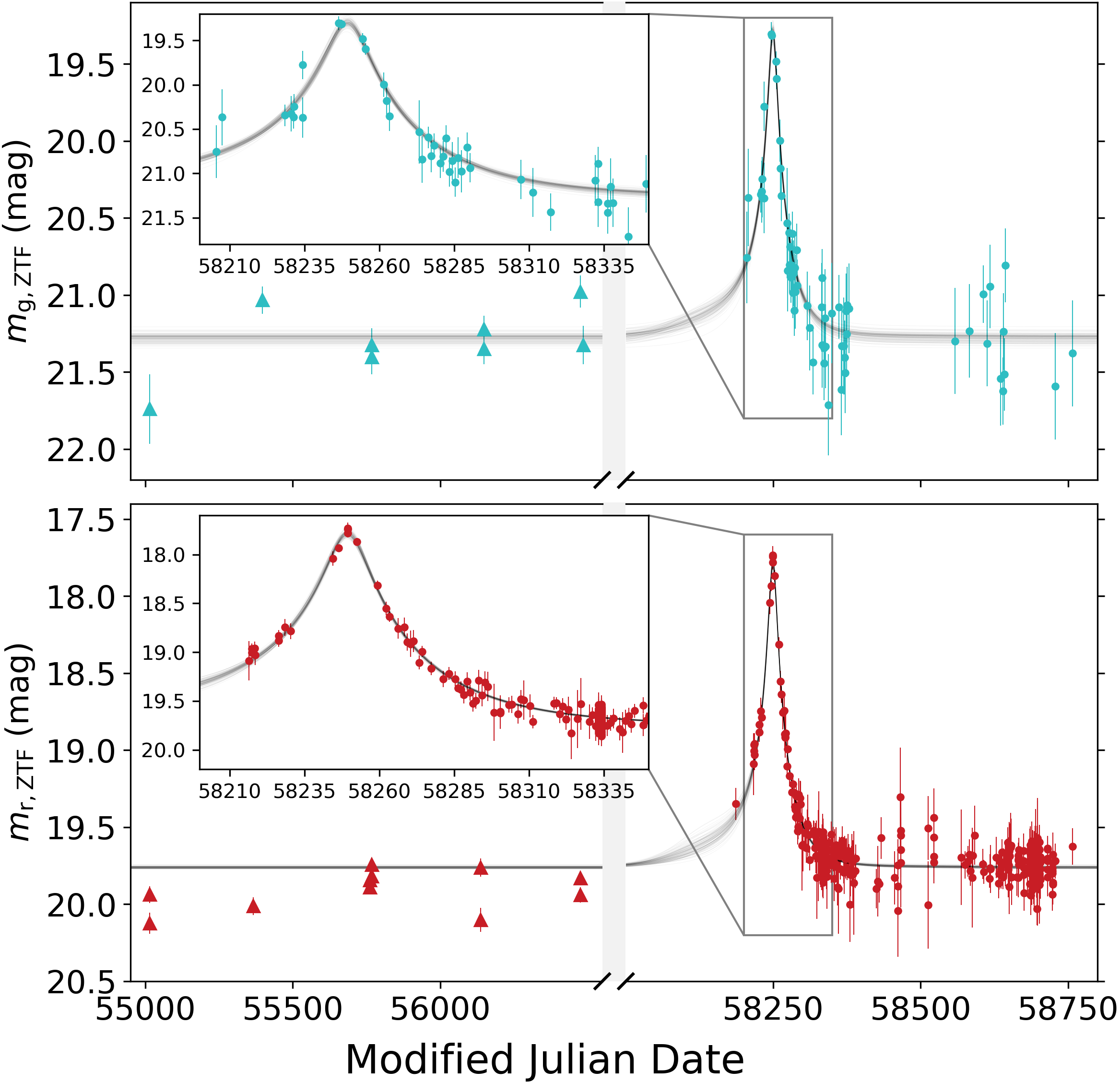}
    \caption{\small{Microlensing photometric lightcurve for ZTF18abhxjmj with ZTF (circles) and PS1 data (triangles), where the PS1 data has been transformed onto the ZTF filter system, in the g-band (top) and the r-band (bottom). 500 draws from the posterior distribution are in light gray for both filters. Note the break in the middle of the plot, as the PS1 data is from 2009 to 2012. The model captures the asymmetry in the rise and fall time due to parallax, but fails to appropriately match the baseline outside of the event with the PS1 r-band data.} \label{fig:ztf1_lightcurve}}
\end{figure}

Figure \ref{fig:ztf1_lightcurve} contains the lightcurves of our example microlensing event which was detected by the ZTF difference imaging alert stream and labelled ZTF18abhxjmj. \citet{Mrz2020} includes this lightcurve in their list of microlensing events detected in the first year of ZTF's Galactic Plane Survey; however we discovered this event independently by our event selection process. ZTF18abhxjmj is located at $(\alpha, \delta) = (284.02920^\circ, 13.15229^\circ)$ or $(\ell, b) = (45.19263^\circ, 4.93715^\circ)$ and began to rise at the start of the ZTF DR1 dataset in March 2018. Pan-STARSS1 (PS1) \citep{PanSTARRS} epochal data shows no previous variability in the years leading up to this event. Measurements in the months after ZTF18abhxjmj also show no variability, although more data at later times would help to better measure the baseline magnitude of the event.

We model ZTF18abhxjmj as a point-source, point-lens event allowing for blending and parallax effects. We transformed PS1 g-band and r-band data into the ZTF filter system to include the data in our fit \citep{Medford2020}, helping to measure the long-duration baseline outside of the event.  Bayesian fitting was performed with nested sampling \citep{Skilling2006} performed by \texttt{PyMultiNest} \citep{Buchner2014}, built on top of \texttt{MultiNest} \citep{Feroz2009}. Our fitter calculates magnifications in a heliocentric reference frame, avoiding the necessity to calculate a parameter reference time $t_{0,par}$. Priors for the Einstein crossing time and microlensing parallax components were taken from one dimensional marginalizations of the microlensing parameters extracted from \texttt{PopSyCLE} simulations pointed at the location of the event, with observational cuts applied to the microlensing populations as described in Section \ref{sec:uLens_estimate}. Following the example of previous work such as \citet{Batista2011}, we apply not generic Galactic priors but priors specific to the mass density, galactic rotation, extinction and consequently the microlensing event rate towards this specific line of sight in the Galaxy. Modelling microlensing events with Bayesian priors derived from \texttt{PopSyCLE} simulations allows for tighter constraints on posteriors than generic priors could otherwise produce.

Figure \ref{fig:ztf1_lightcurve} shows 500 draws from our posterior distributions on top of our ZTF and transformed PS1 data. The model correctly captures the parallax effects near the peak of the event that appear as an asymmetry in the rise and fall time of the lightcurve. The model does not agree with the observed PS1 r-band flux, overestimating this contribution in order to fit the ZTF r-band baseline flux from after the event. The point source estimates from the two $u_0$ solutions in our fit posteriors of ZTF18abhxjmj can be found in Tables \ref{tab:ML_geom_params} and \ref{tab:ML_mag_params}. The event's Einstein crossing time ($t_E$) of 76 days is near the peak of the microlensing distribution for the outer Galaxy line of sight as seen in Figure \ref{fig:einstein_crossing_time}. We note here that our Einstein crossing time ($t_E$), r-band baseline magnitude, r-band blend fraction and parallax components for ZTF18abhxjmj are all in agreement with the parameters found by \citet{Mrz2020} for the same event in their parallax model. Our fit results in different values for $t_0$ and $\pi_E$ which can occur due to the correlation between these variables in the heliocentric reference frame.

\begin{figure}[!htb]
    \centering
    \includegraphics[width=0.45\textwidth, angle=0]{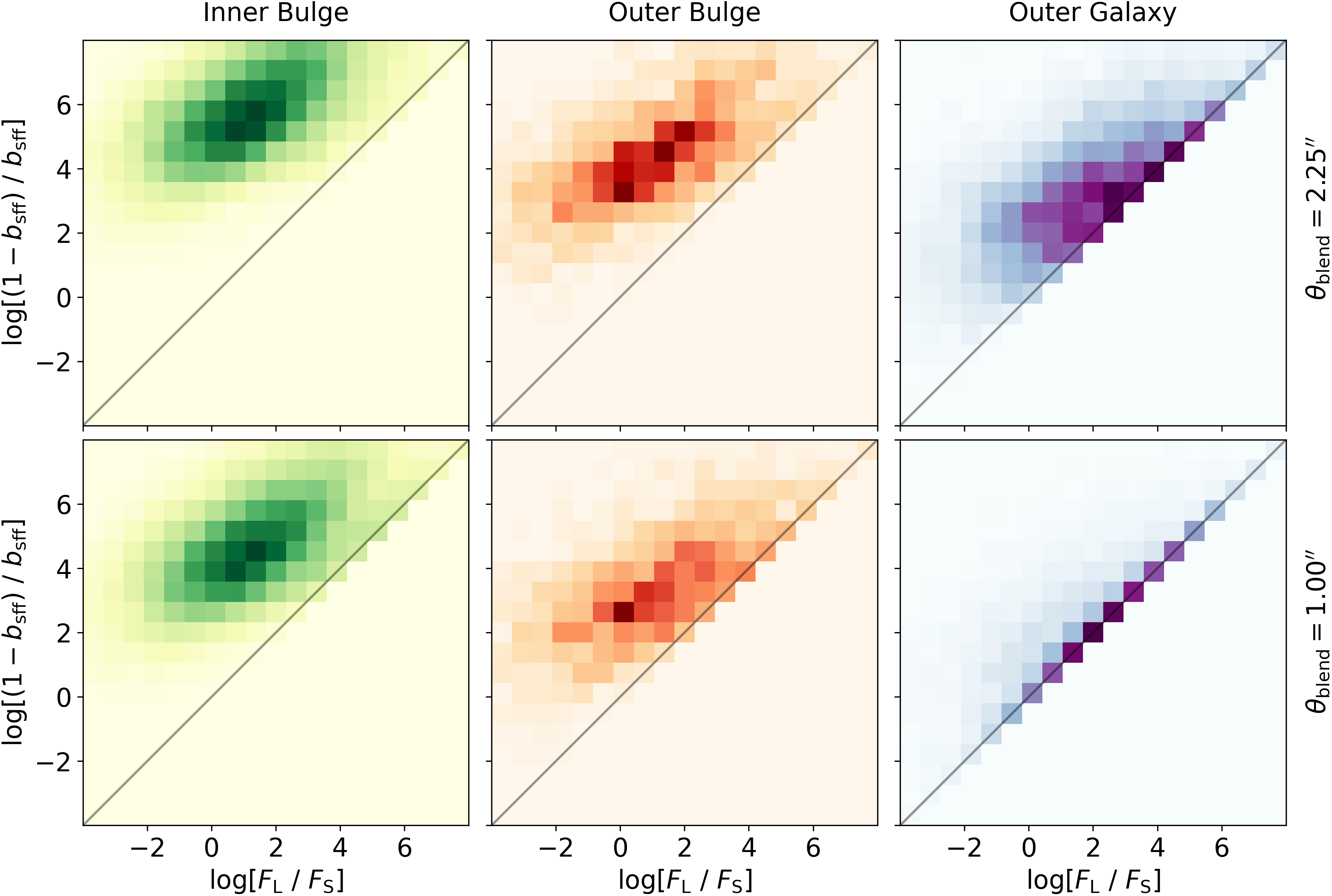}
    \caption{\small{Comparison of the lens-source-flux ratio to it's approximation derived from the source flux fraction (see Equation \ref{eq:bsff_approximator}) in the ZTF r-band across 12 decades for both a larger photometric aperture ($\theta_\text{blend} = 2.25''$) and a smaller aperture ($\theta_\text{blend} = 1.0''$). Events in the outer Galaxy have relatively small contributions to their observable flux from neighboring stars when assuming a smaller aperture, making the source flux fraction approximation valid for almost all events. The presence of neighbor stars is the dominant cause of the spread and offset in the source flux fraction approximation in the remaining observations, making such an approximation invalid in the bulge fields and only partially correct in the larger aperture outer Galaxy field. This approximation enables the conversion from the apparent magnitude of the source to the apparent magnitude of the lens using the source flux fraction in Equation \ref{eq:app_mag_lens}.} \label{fig:f_blend_approximation}}
\end{figure}

Transforming from aperture apparent magnitudes to source and lens apparent magnitudes requires using the source flux fraction, which can often be complicated by the presence of neighbor flux. As discussed in Section \ref{sec:galactic_plane}, very few microlensing events in the outer Galaxy have significant contributions to their flux from neighboring stars when observed with a relatively smaller photometric aperture of $\theta_\text{blend} = 1.0''$. We will assume these optimistic observing conditions because (1) this analysis takes place in the outer Galaxy where there is less confusion due to crowding and (2) the typical seeing on ZTF is around $1.5''$, and therefore an extraction method tuned to these conditions should be able to achieve such a blend radius. Assuming that the presence of neighbor flux is minimal has the convenient consequence of making the measurement of the source flux fraction approximately a measurement of the ratio of source flux to the sum of the flux from both the source and the lens. This approximation can be used to derive the the ratio of flux from the lens and source, or the lens-source-flux ratio, from the source flux fraction as follows:
\begin{align}
	b_\text{sff} &\approx \frac{f_\text{S}}{f_\text{S} + f_\text{L} + 0} \nonumber \\
	\frac{f_\text{L}}{f_\text{S}} &\approx \frac{1 - b_\text{sff}}{b_\text{sff}}. \label{eq:bsff_approximator}
\end{align}
Figure \ref{fig:f_blend_approximation} reveals that this approximation is valid in the outer Galaxy across 12 decades of $b_\text{sff,r}$ values. It is in the Galactic bulge where the abundance of neighbors in the observable aperture makes the source flux fraction approximation an overestimation of the lens-source-flux ratio. Given that our fitter solves for the apparent magnitude of the source, we implement this approximation to calculate the apparent magnitude of the lens in each filter as:
\begin{equation}
    m_\text{L,f} = m_\text{S,f} - 2.5 \log\left(\frac{1 - b_\text{sff,f}}{b_\text{sff,f}}\right)\ ,\ f=\{g, r\}. \label{eq:app_mag_lens}
\end{equation}

\begin{figure}[!htb]
    \centering
    \includegraphics[width=0.45\textwidth, angle=0]{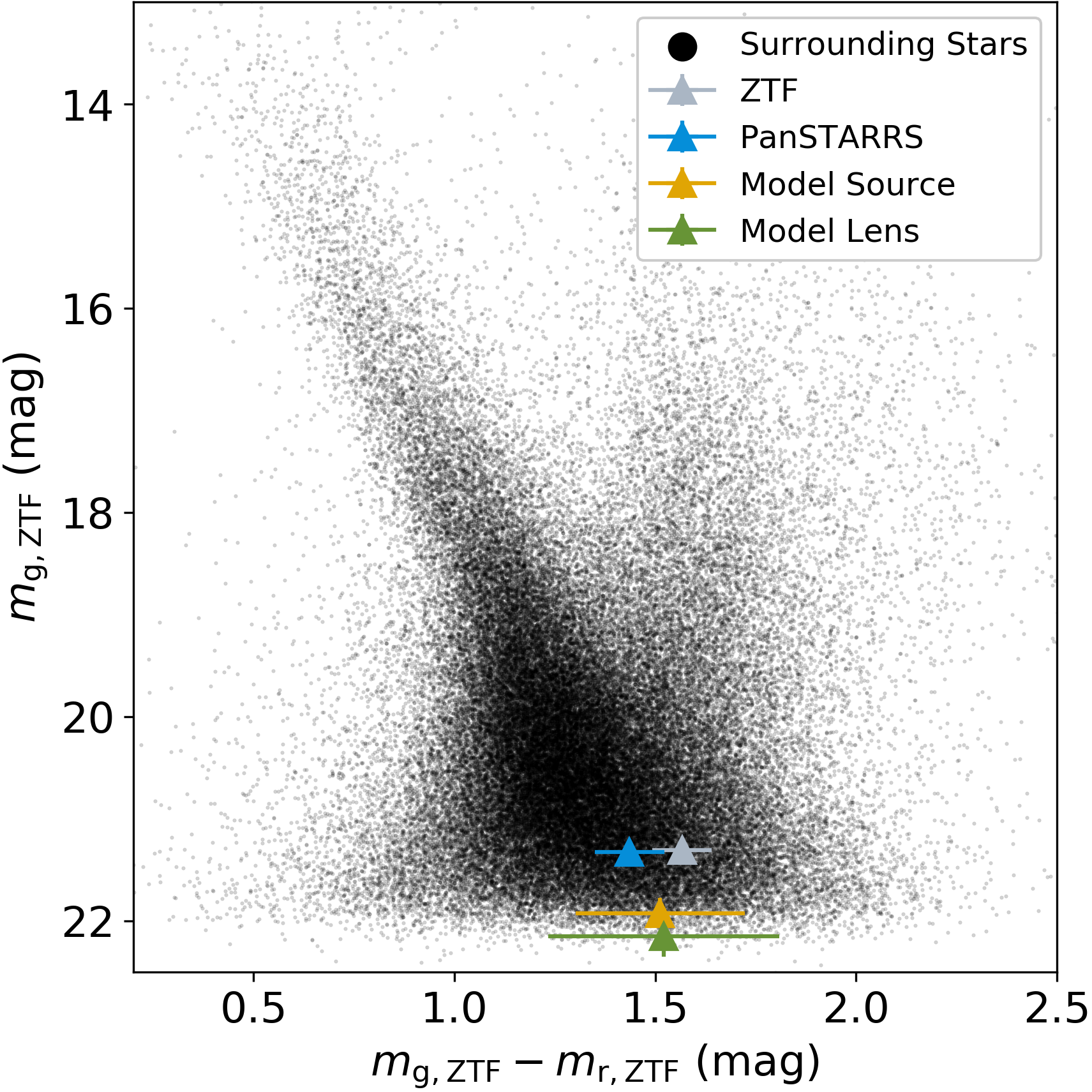}
    \caption{\small{Apparent color-magnitude diagram for 88,914 stars cross-matched in the ZTF g-band and r-band in the 0.77 square degrees surrounding ZTF18abhxjmj. Highlighted are the apparent magnitudes of the event as calculated by the ZTF observations outside of the event (gray), the PS1 observations placed onto the ZTF filter system (blue) and the baseline apparent magnitudes as calculated by the point source point lens model for the source (yellow) and the lens (green). The ZTF measurement is slightly redder than the PS1 measurement, consistent with the mismatched out-of-event flux shown in Figure \ref{fig:ztf1_lightcurve}, but still within the error of the measurement. The g-band source flux fraction of $b_\text{sff,g} = 0.59$ places the source and the lens at nearly the same observable g-band magnitude, while the similar source flux fractions in both filters ($b_\text{sff,g} \approx b_\text{sff,r}$) places the source and the lens at nearly the same observable color.} \label{fig:ztf1_cmd}}
\end{figure}

Figure \ref{fig:ztf1_cmd} presents an apparent color-magnitude diagram of ZTF18abhxjmj (and surrounding stars) that results from folding this approximation into our fitting procedure. The ZTF and PS1 magnitudes and colors are derived from apparent aperture magnitudes taken outside of the microlensing event, while the model magnitudes are derived from the fit. The source and lens appear to have approximately the same apparent color due to their approximately equal source flux fractions ($b_\text{sff,r}\approx b_\text{sff,g}\approx0.59$). The g-band source flux fraction ($b_\text{sff,g}$) is approximately 0.59, meaning that the source and the lens contribute about equally to the apparent g-band brightness. The ZTF color is slightly redder than the PS1 color due to the mismatch in the baseline magnitude in the lightcurve. The model attributes a color to the source and lens between these two values with appropriately larger errors bars, reflecting this discrepancy.

\begin{figure}[!htb]
    \centering
    \includegraphics[width=0.45\textwidth, angle=0]{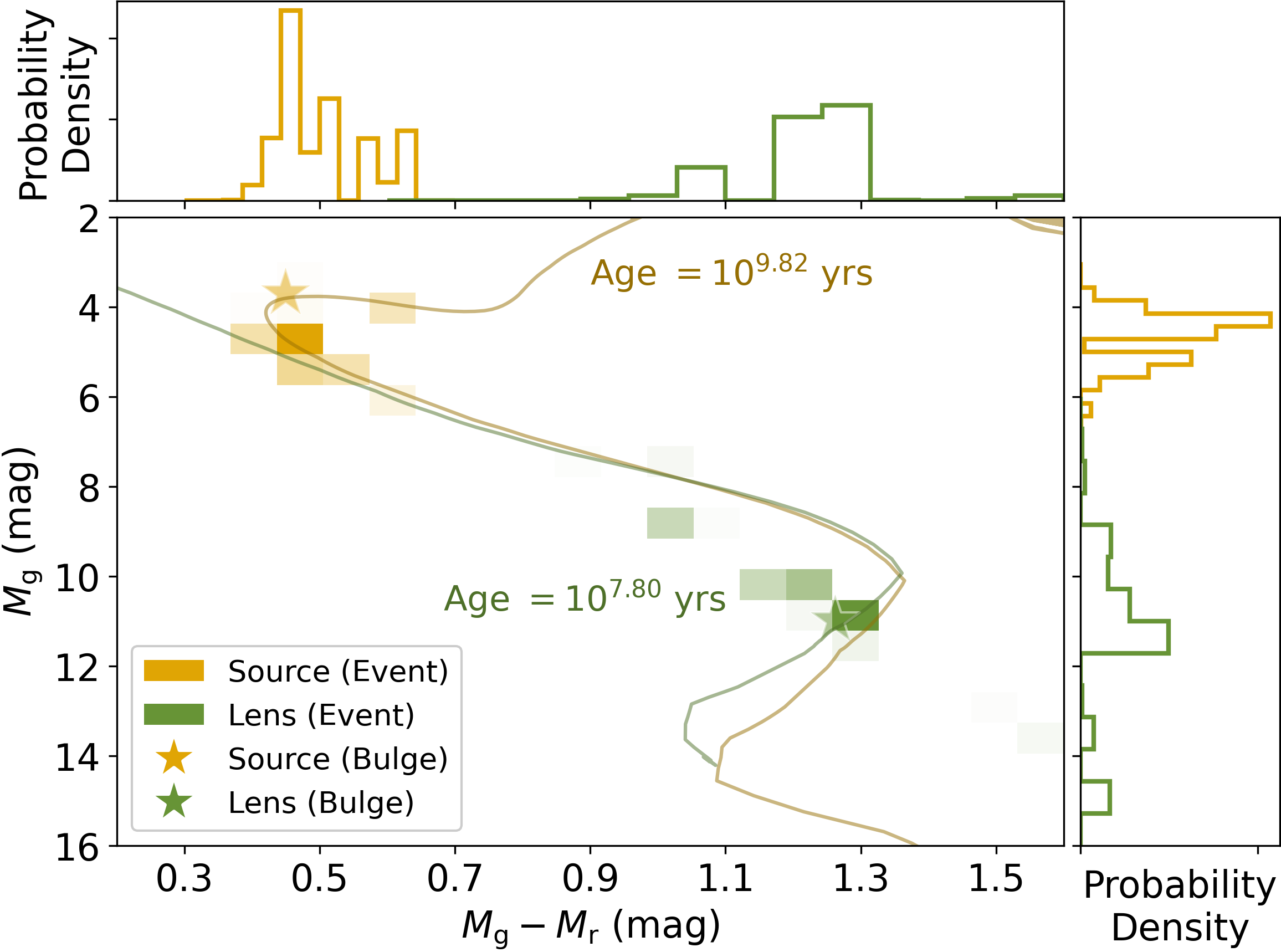}
    \caption{\small{Absolute color-magnitude diagram for the source (yellow) and lens (green) derived from combining Bayesian modeling and \texttt{PopSyCLE} simulations, with histograms on both axes showing the marginalized distributions of the parameters. Isochrones generated by \texttt{PyPopStar} have been drawn to approximate the source and lens ages to be $10^{9.82}$ years and $10^{7.8}$ years respectively. Point estimates for the source and lens calculated using \texttt{PopSyCLE} catalogs generated in the outer bulge (stars) find a slightly brighter source due to the additional extinction in the Galactic bulge. These estimations are highly sensitive to the systematic errors discussed throughout Section \ref{sec:uLens_discovery}.} \label{fig:ztf1_hr}}
\end{figure}

Calculating the absolute magnitudes and stellar types of the source and lens from their apparent magnitudes requires knowing their distances and extinctions. 
As discussed in Section \ref{sec:galactic_plane}, \texttt{PopSyCLE} produces distributions of distances and extinctions for microlensing events along a specific line of sight. 
We generated \texttt{PopSyCLE} simulations at the location of ZTF18abhxjmj and applied observational cuts to the event catalogs that simulated ZTF observing conditions. 
Samples were drawn from these trimmed catalogs, weighted by the event model's Bayesian posteriors for baseline magnitude and source flux fraction in g-band and r-band. 
Figure \ref{fig:ztf1_hr} shows the absolute color-magnitude diagram of the samples that resulted from this procedure.

The source of ZTF18abhxjmj has an absolute magnitude in the g-band of $M_\text{S,g}=4.6 \pm 0.6$ and an absolute color of $M_\text{S,g-r}=0.49 \pm 0.07$, while the lens has an absolute g-band magnitude of $M_\text{L,g}=11.1 \pm 2.6$ and an absolute color of $M_\text{L,g-r}=1.3 \pm 0.36$. We matched these source and lens absolute magnitudes to absolute magnitudes of stars generated in synthetic clusters with \texttt{PyPopStar} \citep{Hosek2020}, a python package that generates single-age, single-metallicity populations from user specified initial mass functions, stellar evolution models, and stellar atmospheres. The source approximately resembles a 1.04 solar mass G-star in a $10^{9.82}$ year old cluster, and the lens approximately resembles a 0.39 solar mass M-dwarf in a $10^{7.8}$ year old cluster. Systematic errors in the Galactic model implemented in \texttt{PopSyCLE} significantly contribute to the uncertainty in these conclusions but are not captured by our stated errors.

We have included in Figure \ref{fig:ztf1_hr} the source and lens absolute magnitudes that would have been calculated if a simulated catalog from the outer bulge was used instead of one produced along the target's line of sight. Microlensing source and lenses towards the bulge are, on average, at closer distances and are behind more magnitudes of extinction. These two facts have opposite effects on the estimate of the source's absolute magnitude. The additional extinction pushes the source star's probability to a smaller absolute magnitude in order for the source or lens to appear at the apparent magnitude determined by the Bayesian fit, with the closer distance having the opposite effect. The results of these two competing effects can be resolved with \texttt{PopSyCLE} simulations 
at the location of each microlensing event that ZTF discovers modelled after this fitting procedure.

This example analysis demonstrates how data from ZTF and simulations from \texttt{PopSyCLE} can be combined to fit microlensing models and estimate stellar types of microlensing sources and lenses. The results of this particular analysis are not exceptional as M-dwarfs are extremely common throughout the Galaxy and are often found to be lenses of microlensing events, although this method could be used to find more exotic lenses such as free-floating planets and black holes. We have outlined the steps of this analysis to illustrate how probabilistic priors for a specific event can be quickly generated through modelling microlensing populations toward a particular line of sight.

\section{Discussion} \label{sec:conclusion}

The Zwicky Transient Facility and its surveys are an excellent opportunity to discover microlensing events. 
We find that ZTF will observe $\sim$1100 events in three years of observing, with $\sim$500 events occurring outside of the Galactic bulge in the outer Galaxy ($\ell \geq 10^\circ$). 
This total can be increased to $\sim$2400 events ($\sim$1300 events in the outer Galaxy) by extending ZTF operations to five years and executing a post-processing image co-addition pipeline. The event rate of microlensing is proportional to the number of observed luminous sources. 
While ZTF's single image limiting magnitude is not as deep as other optical surveys, it's massive 49 deg$^2$ camera is able to cover the entire northern sky every three nights in multiple filters. 
The decrease in microlensing event rate outside of the Galactic bulge that discourages other microlensing surveys is compensated for by the billions of stars observed within this large footprint. Observing in the outer Galaxy almost doubles the total number of microlensing events that ZTF will observe.

Microlensing events can be discovered in ZTF by searching through the epochal photometric catalogs present in the public data releases described in Section \ref{sec:ZTF}. 
These catalogs contain observations in multiple filters that allow for confirming a potential microlensing event through its achromaticity.
ZTF also generates subtraction images for all of its exposures and serves a real-time alert stream of transient detections found on these difference images. 
Filters could be developed that search for microlensing events on a nightly basis \citep{PriceWhelan2014, Godines2019}, generating a list of candidates that could trigger photometric or astrometric followup. 
This would be particularly helpful in attempting to detect exoplanets through microlensing, which requires triggering higher cadence followup near the photometric peak of the event, as well as discovering black holes lenses which requires astrometric follow-up.

Microlensing detections made outside of the Galactic plane will be extremely rare due to the decrease in luminous stellar sources. 
Galaxies begin to be the dominant luminous sources in these fields and the distance ratio of luminous sources and massive lenses does not result in observable microlensing events. 
Galaxies are far away and microlensing is maximized when the lens is halfway between the source and the observer, so we therefore cannot hope to observe any microlensing events where galaxies are the luminous source. 
However this challenge can be inverted to provide an interesting opportunity. 
There is a possibility that primordial black holes (PBHs) significantly contribute to dark matter and could be observed through microlensing.
Previous work suggest that the dark matter mass fraction contributed to by PBHs could be constrained through an effect on the shape of the Einstein crossing time distribution \citep{Green2016, Green2017, Niikura_2019, Lu2019}. 
Given the lack of observable microlensing events outside of the Galactic plane, and the isotropic distribution of dark matter, any microlensing detections made outside the plane could place constraints on the PBH dark matter fraction. 
The likelihood that a microlensing event is caused by a PBH lens relative to a stellar lens increases when observing outside the Galactic plane.
A ZTF microlensing survey would be one of the only microlensing surveys conducted that includes observing in these fields, making it one of the few surveys that could make this measurement.
\begin{figure}[!htb]
    \centering
    \includegraphics[width=0.45\textwidth, angle=0]{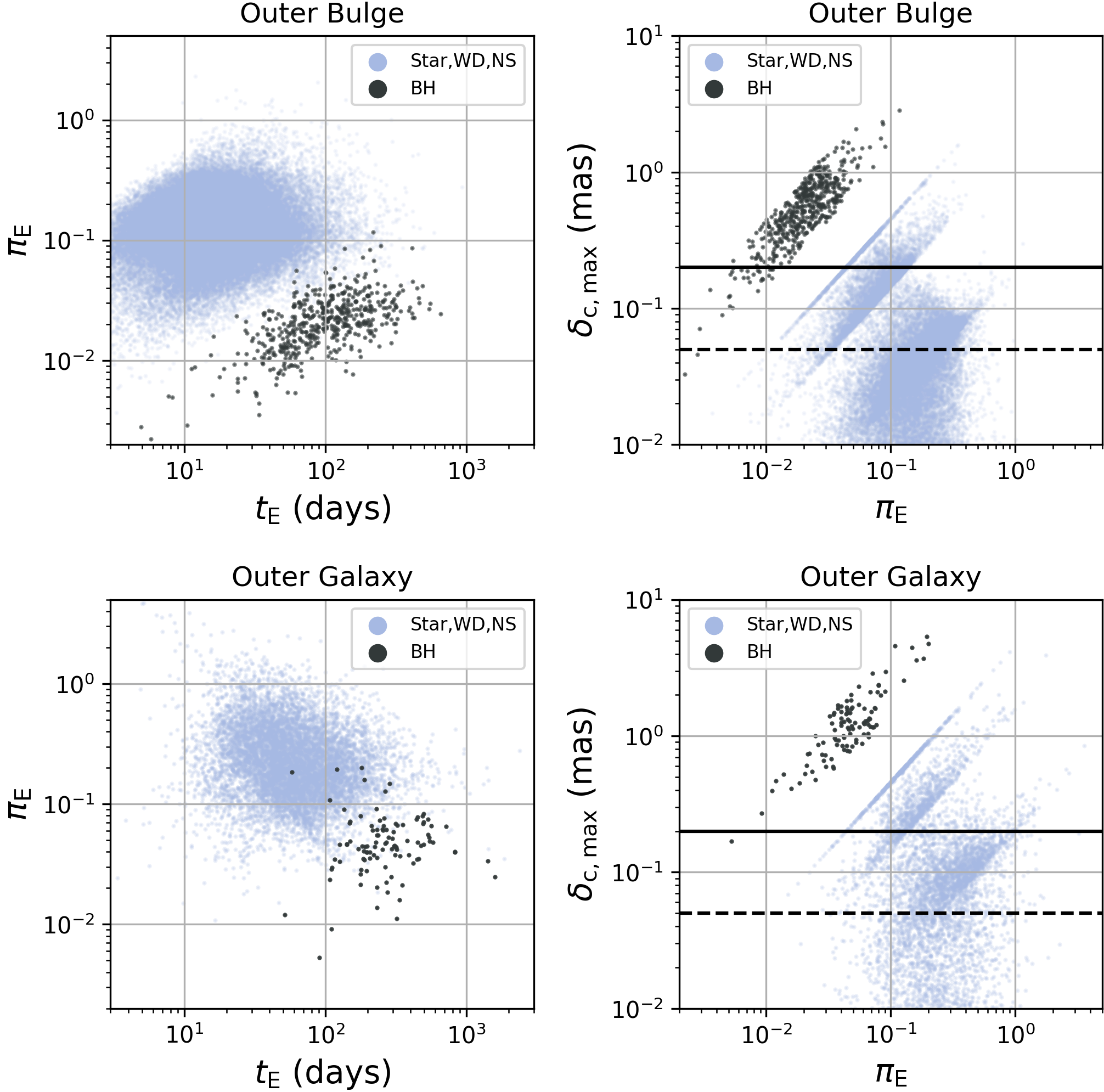}
    \caption{\small{Microlensing parallax $\pi_E$ vs. Einstein crossing time (top) and maximum astrometric shift $\delta_\text{c,max}$ vs. mircolensing parallax $\pi_E$ broken out by astrophysical type of lens for the outer bulge (top) and outer Galaxy (bottom) fields. \texttt{PopSyCLE} simulations reveal that black hole microlensing lenses are distinct from stars, white drawfs and neutron stars in these two spaces \citep{Lam2020}. Both the Einstein crossing times and microlensing parallaxes increase when measured in the outer Galaxy (bottom) as compared to the outer bulge (top), making it easier to constrain black holes in this plane. The maximum astrometric shift for black holes increases to a decade above the detection limit of the Keck laser guide star adaptive optics system (solid) and almost two decades above the anticipated limits of the Nancy Grace Roman Space Telescope or the Thirty Meter Telescope (dashed). Introducing observational cuts not present in these figures reduces the total number of events but maintains the same trends.} \label{fig:delta_c_max}}
\end{figure}
There may also be advantages in looking for black holes as microlensing lenses in the outer Galaxy as compared to the Galactic bulge. Detecting a black hole through microlensing requires weighing the mass of the lens despite the lens mass' degeneracy with microlensing parallax when using photometric data. This degeneracy can be avoided by astrometric measurement which can determine the mass of the lens directly. As discussed in Section \ref{sec:einstein_radius}, outer Galaxy microlensing events have larger Einstein radii and therefore have an astrometric signature that is easier to detect. \citet{Lam2020} outline how black hole lenses have significantly larger maximum astrometric shifts, longer Einstein crossing times and less microlensing parallax than star, white dwarfs or neutron stars lenses with \texttt{PopSyCLE} simulations. Figure \ref{fig:delta_c_max} replicates Figures 12 and 13 from \citet{Lam2020} in our outer bulge and outer Galaxy fields. All events in the outer Galaxy sample occur at longer Einstein crossing times and with larger microlensing parallaxes, making them easier to measure and therefore distinguish black hole lenses. The maximum astrometric shift is significantly larger, averaging almost an order of magnitude above the 0.2 miliarcseconds that the Keck laser guide star adaptive optics system is capable of measuring \citep{Lu2016} and maxing out at over 5 miliarcseconds. Decreased stellar densities in the outer Galaxy will present a challenge to making this measurement and requires the Hubble Space Telescope or wide field adaptive optics such as an upgraded Gemini North adaptive optics system if measured from the ground. Future space instruments such as the James Webb Space Telescope \citep{Gardner2006} or the Nancy Grace Roman Space Telescope \citep{Spergel2013} will be more than capable of detecting black holes using this technique.

There are also challenges that arise when attempting to use ZTF to make microlensing measurements. 
ZTF's photometric precision of $\sim$0.1 magnitudes at a limiting magnitude of $m_\text{lim} < 21$ \citep{Masci2018} can make it difficult to detect events with a large impact parameter or small maximum amplification. 
These events will be difficult to distinguish from background noise or variability of faint stars. ZTF is located in the Northern hemisphere, limiting exposure to the Galactic plane to select summer months of the year, reducing the total number of observable short duration events. ZTF is also a collaboration with many priorities both Galactic and extra-Galactic resulting in decisions on survey design, cadence and scientific goals that are not necessarily optimized for microlensing.

Future synoptic surveys such as the Rubin Observatory Legacy Survey of Space and Time (LSST) could continually monitor billions of stars across the Milky Way for many years, providing opportunities to learn about galactic structure, stellar populations and possibly even dark matter through photometric microlensing. The massive footprints of surveys such as ZTF and LSST unlock the potential to observe thousands of microlensing events across the entire Galactic plane and possibly even off the plane, expanding beyond the scope of microlensing surveys to date that have been pointed at the Galactic bulge and other nearby galaxies. Combining these datasets with sophisticated microlensing modelling software can result in improvements to stellar categorization and population statistics that would otherwise be out of reach for these photometric surveys. 

\begin{deluxetable*}{ccccccccccc}
\tablecolumns{11}
\tablewidth{0pt}
\tablecaption{Microlensing Parameters of ZTF18abhxjmj}
\tablehead{
\colhead{$\ell$} & \colhead{$b$} & \colhead{$t_0$} & \colhead{$t_E$} & \colhead{$u_0$} & \colhead{$m_\text{S,g}$} & \colhead{$b_\text{sff,g}$} & \colhead{$m_\text{S,r}$} & \colhead{$b_\text{sff,r}$} & \colhead{$\pi_\text{E,N}$} & \colhead{$\pi_\text{E,E}$} \\
\colhead{deg.} & \colhead{deg.} & \colhead{MJD} & \colhead{days} & \colhead{-} & \colhead{mag} & \colhead{-} & \colhead{mag} & \colhead{-} & \colhead{-} & \colhead{-}}
\startdata
284.02916 & 13.15228 & 58229.9 & 76.7 & 0.14 & 21.84 & 0.592 & 20.33 & 0.593 & 0.187 & 0.257 \\
& & \plusminus{4.0}{4.1} & \plusminus{8.7}{8.8} & \plusminus{0.04}{0.03} & \plusminus{0.17}{0.15} & \plusminus{0.077}{0.102} & \plusminus{0.17}{0.15} & \plusminus{0.078}{0.097} & \plusminus{0.054}{0.040} & \plusminus{0.050}{0.036} \\
& & 58227.1 & 75.8 & -0.05 & 21.85 & 0.591 & 20.33 & 0.589 & 0.198 & 0.241 \\
& & \plusminus{3.7}{4.0} & \plusminus{8.0}{6.6} & \plusminus{0.05}{0.05} & \plusminus{0.16}{0.12} & \plusminus{0.065}{0.096} & \plusminus{0.16}{0.11} & \plusminus{0.056}{0.094} & \plusminus{0.047}{0.036} & \plusminus{0.051}{0.036} \\
\enddata
\tablecomments{\small{The microlensing parameters of the median best-fit point-source point-lens microlensing model of ZTF18abhxjmj, including the time of maximum heliocentric amplification ($t_0$), Einstein crossing time ($t_E$), minimum source-lens separation in units of the Einstein radius ($u_0$), baseline magnitudes for the source in g-band and r-band ($m_\text{S,g}, m_\text{S,r}$), source-flux-fractions in g-band and r-band ($b_\text{sff,g}, b_\text{sff,r}$) and the two components of the microlensing parallax ($\pi_\text{E,N}, \pi_\text{E,E}$). We find an Einstein crossing time of 76 days in our two $u_0$ solutions and a blend fraction in both g-band and r-band around 0.59. These values indicate that the flux in the aperture is about equally split between the source and the lens in both filters. The visible parallax in the lightcurve appears in the fit, confirmed by significant components of $\pi_E$.}}
\label{tab:ML_geom_params}
\end{deluxetable*}

\begin{deluxetable*}{c||rrr}
\tablecolumns{7}
\tablewidth{10pt}
\tablecaption{Model Magnitudes of ZTF18abhxjmj}
\tablehead{& \colhead{$M_\text{g}$} & \colhead{$M_\text{r}$} & \colhead{$M_\text{g-r}$}}
\startdata
Lens & 11.12 $\pm$ 2.64 & 9.84 $\pm$ 2.28 & 1.28 $\pm$ 0.36 \\
Source & 4.58 $\pm$ 0.59 & 4.08 $\pm$ 0.57 & 0.49 $\pm$ 0.07 \\
\enddata
\tablecomments{\small{The absolute magnitudes ($M_\text{g},\ M_\text{r}$), and absolute color ($M_\text{g-r}$) of the point-source point-lens microlensing model of ZTF18abhxjmj. The absolute magnitudes are calculated by drawing samples from the \texttt{PopSyCLE} simulations generated at the location of the event weighted by the posteriors of our Bayesian fit. The errors on these measurements do not include systematics from \texttt{PopSyCLE}'s Galactic model.}}
\label{tab:ML_mag_params}
\end{deluxetable*}

\vspace{0.1in}
\noindent
{\it Acknowledgements:} This work was done with support from the University of California Office of the President for the UC Laboratory Fees Research Program In-Residence Graduate Fellowship (Grant ID: LGF-19-600357). 
Part of this work was performed under the auspices of the U.S. Department of Energy by Lawrence Livermore National Laboratory under Contract DE-AC52-07NA27344 and was supported by the LLNL-LDRD Program under Project Number 17-ERD-120. J.R.L. and C.L. acknowledge funding from the National Science Foundation (AST-1909641).
Authors would also like to thank Danny Goldstein and Matt Hosek for their support both technical and otherwise.

This document was prepared as an account of work sponsored by an agency of the United States government. Neither the United States government nor Lawrence Livermore National Security, LLC, nor any of their employees makes any warranty, expressed or implied, or assumes any legal liability or responsibility for the accuracy, completeness, or usefulness of any information, apparatus, product, or process disclosed, or represents that its use would not infringe privately owned rights. Reference herein to any specific commercial product, process, or service by trade name, trademark, manufacturer, or otherwise does not necessarily constitute or imply its endorsement, recommendation, or favoring by the United States government or Lawrence Livermore National Security, LLC. The views and opinions of authors expressed herein do not necessarily state or reflect those of the United States government or Lawrence Livermore National Security, LLC, and shall not be used for advertising or product endorsement purposes. Lastly we thank the lawyers of Lawrence Livermore National Laboratory for their contributions to this acknowledgement section.

\bibliography{references}

\end{document}